\documentclass[%
 reprint,
 amsmath,amssymb,
 aps,
 prb,
]{revtex4-1}

\usepackage{graphicx}
\usepackage{amssymb}
\usepackage{amsfonts}
\usepackage{amsmath}
\usepackage{amsbsy}
\usepackage{natbib}
\usepackage{comment}

\usepackage{color}
\usepackage{float}

\renewcommand{\vec}[1]{\mbox{\boldmath$\mathrm{#1}$}}
\newcommand{\mycite}[1]{\scalebox{1.5}[1.5]{\raisebox{-0.85ex}{\cite{#1}}}}

\begin{document}

%\preprint{APS/123-QED}

\title{High-fidelity magnonic gates for surface spin waves}

\author{Xi-guang Wang$^{1}$, L. Chotorlishvili$^2$, Guang-hua Guo$^{1}$, J. Berakdar$^2$}

\address{$^1$ School of Physics and Electronics, Central South University, Changsha 410083, China \\
$^2$ Institut f\"ur Physik, Martin-Luther Universit\"at Halle-Wittenberg, 06099 Halle/Saale, Germany
}

\date{\today}% It is always \today, today,

\begin{abstract}
We study the propagation of surface spin waves in two wave guides coupled through the dipole-dipole interaction. Essential for the observations made here is the magneto-electric coupling between the spin waves and the effective ferroelectric polarization. This allows an external electric field to act on spin waves and to modify the band gaps of magnonic excitations in individual layers. By an on/off switching of the electric field and/or varying its strength or  direction  with respect to the  equilibrium magnetization, it is possible to permit or ban the propagation of the spin waves in  selected waveguide. We propose experimentally feasible nanoscale device operating as a high fidelity surface wave magnonic gate.
\end{abstract}

\maketitle
 Magnetostatic surface waves (MSSWs) \cite{Murakami, Zhang, WangZhangWang, Freedman, Matsumoto, Iacocca,Chumak,Khitun,Demokritov} are key elements for
 magnonic-based information transfer and processing.  For realizing efficient magnonic circuits, reliable coupled MSSW waveguides are essential, as they serve as information channel or signal splitter \cite{e1701517,Sadovnikov192406}.
In the present work, we propose a scheme of an electric-field controlled coupled surface MSSWs. Microscopically, the  mechanism is driven by spin-orbit coupling that leads to the emergence of a spin-driven electric polarization coupled to spin-noncollinearity, as discussed for example  for Yttrium iron garnet (YIG) \cite{Flatte, Vignale}. The mechanism can be viewed as a dynamical Dzyaloshinskii Moriya (DM) coupling, and will be named as such hereafter.
 Thus, an external electric field that affects the electric polarization steers indirectly the spin order. In this work we will focus on YIG as it has low magnetic damping constant,  meaning magnetic losses are negligible on relatively  short time scale \cite{Serga264002}. The remarkable property of MSSWs is their chiral nature. The chiral character of MSSWs is a purely dynamical effect related to the off-diagonal part of the dynamic dipole-dipole interaction.  The dynamically induced chirality leads at least to two exciting phenomena:  The non-reciprocal localization of MSSWs propagating perpendicular to the magnetization direction in an in-plane magnetized thin film, and  b) the ban on the backscattering, i.e., robust spin-wave modes with the frequency located inside the volume modes (VM) gap, see \cite{Mohseni, Uchida, Slavin, Kostylev}, and references therein.
In the present work, we study the influence of the dynamical DM interaction on the backscattering-immune frequency region in a single MSSW and coupled wave guides which are coupled via the dipole-dipole interaction. We show that the non-reciprocal localization of MSSWs in the coupled wave guides leads to  non-reciprocal SW dispersion. Through the DM interaction term, an electric field influences the MSSWs dispersion relations and changes the backscattering-immune frequency region.  Furthermore, we study the transfer of MSSWs between two coupled wave guides.

\section{Theoretical model}
For studying the effect of an applied external electric field on the excitation and propagation of  MSSWs in a thin YIG film, we start from the Landau-Lifshitz-Gilbert (LLG) equation supplemented by the magnetoelectric coupling term \cite{jap073903,prb064426}
\begin{equation}
\displaystyle \frac{\partial \vec{M}}{\partial t} = - \gamma \vec{M} \times \bigg(\vec{H}_{\mathrm{eff}} - \frac{1}{\mu_0 M_s} \frac{\delta E_{\mathrm{elec}}}{\delta \vec{m}} \bigg)+ \frac{\alpha}{M_{s}} \vec{M} \times \frac{\partial \vec{M}}{\partial t}.
\label{LLG}
\end{equation}
Here $ \vec{M} = M_s \vec{m} $, $ M_s $ is the saturation magnetization, $ \gamma $ is the gyromagnetic ratio, $ \mu_0 = 4 \pi \times 10^{-7} $ N/A$ ^2 $ is the permeability, and $ \alpha $ is the phenomenological Gilbert damping constant. The effective field $ \vec{H}_{\mathrm{eff}} = \frac{2 A_{ex}}{\mu_0 M_s} \Delta \vec{m} + \vec{H}_{\mathrm{demag}} + H_0 \vec{y} $ consists of the exchange field, the demagnetization field, and the external magnetic field applied along the $ \vec{y} $ axis. $ A_{ex} $ is the exchange constant. The magnetoelectric coupling term $ E_{\mathrm{elec}} = - \vec{E} \cdot \vec{P} $ describes the coupling between the spin-driven  ferroelectric polarization $ \vec{P} $ and applied external electric field $ \vec{E} $. This ferroelectric polarization of YIG in the continuous limit has the form $ \vec{P} = c_E [(\vec{m} \cdot \nabla) \vec{m} - \vec{m}(\nabla \cdot \vec{m})] $  which makes clear the similarity to  the  Dzyaloshinskii Moriya interaction (DMI).
Microscopically, the macroscopic effective electric polarization is magnetically driven. The  dynamic spin non-collinearity results in an effective electric  polarization $\vec{P}$. The driving mechanism is effectively a dynamic   DMI  with a DMI vector  $ \vec{D} = -J\frac{ea}{E_{SO}}\vec{E}\times\vec{e}_{n,n+1}$, as discussed for example in Ref. 
[\mycite{Vignale}]. Here, $E_{SO}=\hbar^{2}/2m_{e}\lambda^{2}$, $ J $ is the exchange coefficient, $m_{e}$ is the mass of the electron, $e$ is the electron charge, $\lambda$ is the spin-orbital coupling constant,  $\vec{e}_{n,n+1}$ is the unit vector connecting the magnetic ions. For the  parameters $ |\vec{E}| = 3.4 $ MV/cm,  $E_{SO}=$3.4 eV ($ \lambda = 1 \AA $) and $a=12\cdot 10^{-10}$ m, the effective DMI constant scales with the exchange constant $D/J = 0.12$. This estimation is in line with experimental observations in Ref. [\mycite {Flatte}].
The demagnetization field in presence of the magnetic dipole-dipole interaction reads
\begin{equation}
\displaystyle \vec{H}_{\mathrm{demag}}(\vec{r}) = -\frac{M_s}{4 \pi} \int_{V} \nabla \nabla'\frac{1}{|\vec{r} - \vec{r'}|}\vec{m}(\vec{r'})d\vec{r}' .
\label{demag}
\end{equation}

In numerical calculations, we adopt the material parameters for YIG: $ M_s = 1.4 \times 10^5 $ A/m, $ A_{ex} = 3 \times 10^{-12} $ J/m, the damping constant $ \alpha = 0.001 $ and magneto electric coupling $ c_E = 0.9 $ pC/m. The magnetic field of the amplitude $ H_0 = 4 \times 10^5 $ A/m is applied along the $ y $ axis (cf. for example Fig.\ref{single-amp}).

\section{spin wave dispersion}

\begin{figure}
    \includegraphics[width=0.48\textwidth]{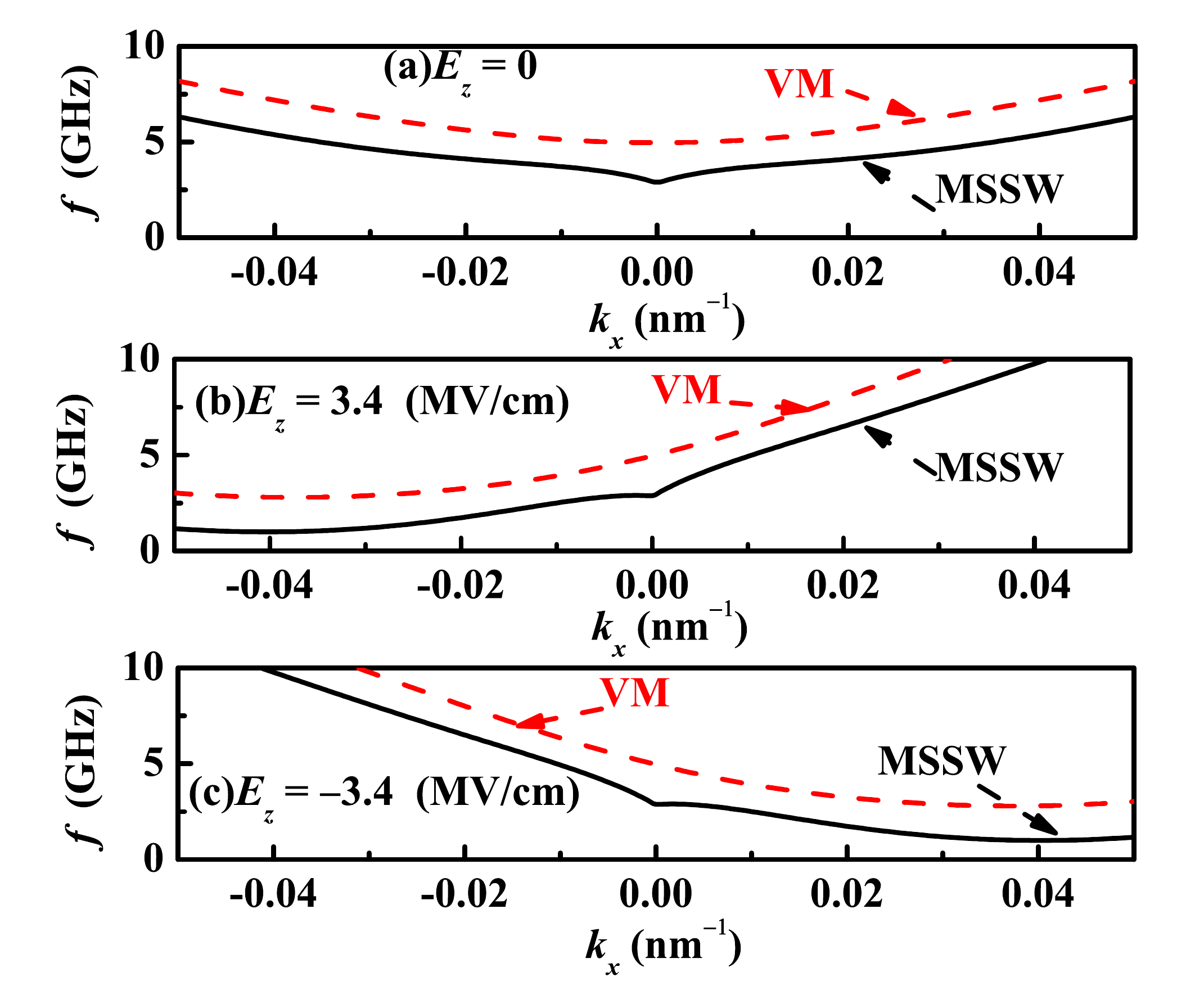}
    \caption{\label{sim-dispersion} The spin-wave dispersion relation simulated for a single film in the presence of an electric field $\vec E_z$ along the $z$ direction with the amplitude: $ E_z = 0 $ (a), $ E_z = 3.4 $ MV/cm (b) and $ E_z = -3.4 $ MV/cm (c). The MSSWs and volume modes (VM) are shown by the black solid lines and red dashed lines, respectively. }
\end{figure}

\begin{figure}
	\includegraphics[width=0.48\textwidth]{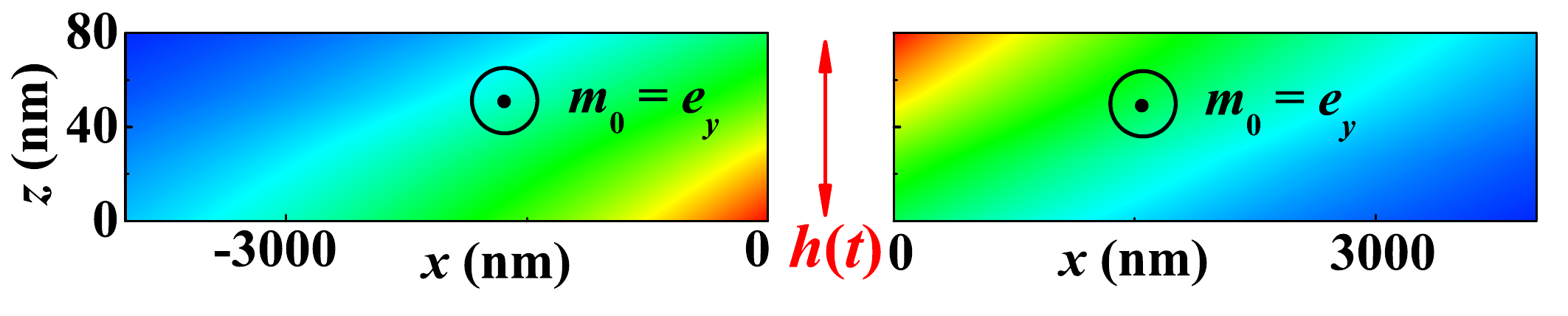}
	\caption{\label{single-amp} Spatial profiles of MSSWs amplitudes propagating in the single film without electric field ($ E_z = 0 $). MSSW is excited in $ x= 0 $ by the local microwave field $ h(t) $. The frequency of the field is equal to 5 GHz.}
\end{figure}

\begin{figure}
    \includegraphics[width=0.48\textwidth]{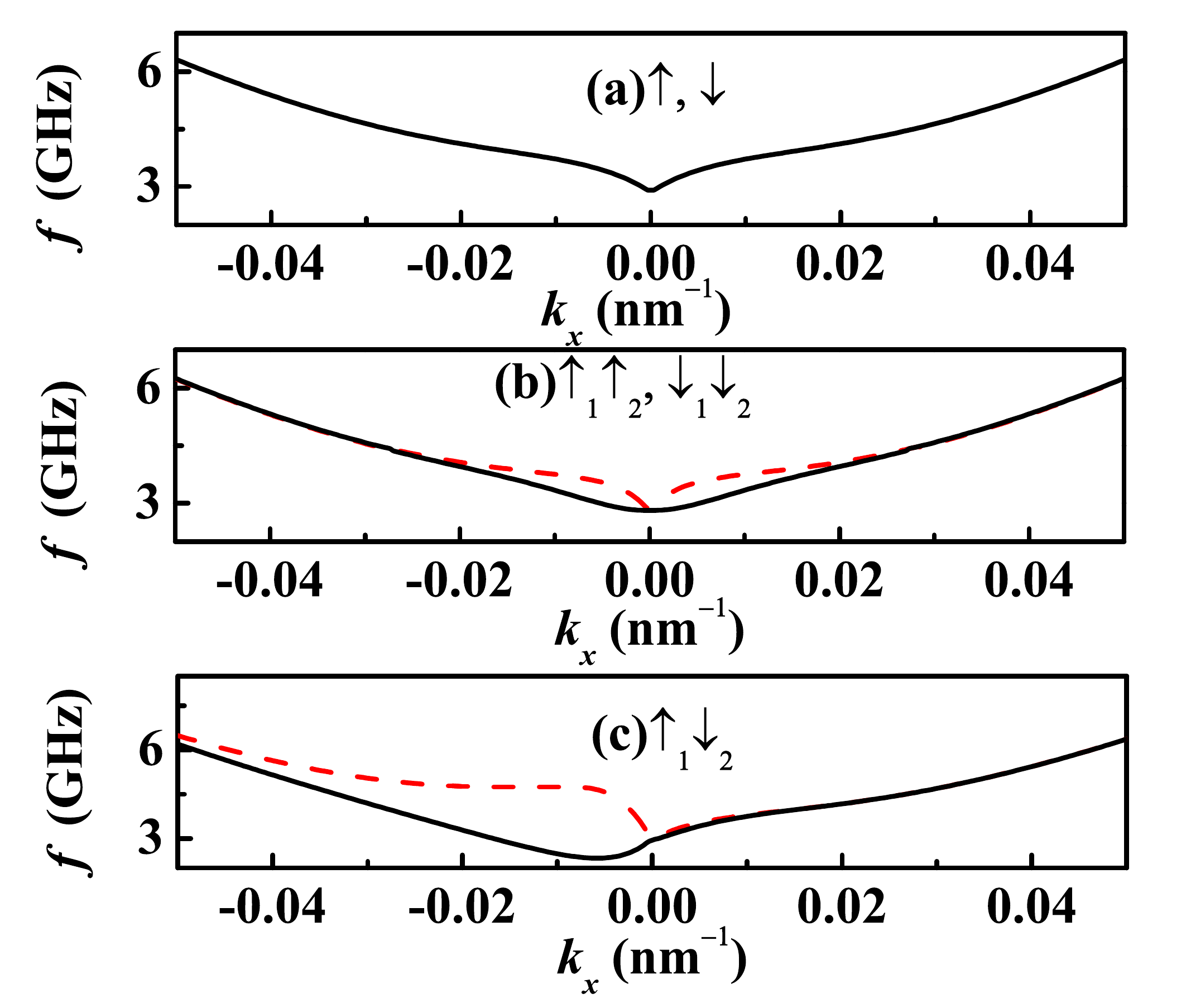}
    \caption{\label{sim-dispersion-coupled} The simulated spin-wave dispersion relation in (a) single film (the equilibrium magnetization is oriented along the $ + \vec{e}_y $ ($ \uparrow $) or $ -\vec{e}_y $  ($ \downarrow $)) direction. (b) Same for two  films with parallel  equilibrium magnetization  ($ \vec{m}_{0,1} $ and $ \vec{m}_{0,2} $ with both   aligned along $ \uparrow \uparrow$ or $ \downarrow \downarrow$).  (c) Two films with anti-parallel  equilibrium magnetization ($ \vec{m}_{0,1} = \uparrow $ and $ \vec{m}_{0,2} = \downarrow $). If $ \vec{m}_{0,1} = \downarrow $ and $ \vec{m}_{0,2} = \uparrow $ i.e, in the case of  anti-parallel equilibrium magnetization of the films, the dispersion relation is asymmetric with respect to the point $ k_x = 0 $. Black solid and red dashed curves for the coupled layers represent acoustic and optic surface wave modes, respectively.}
\end{figure}

\textbf{\begin{figure}
        \includegraphics[width=0.48\textwidth]{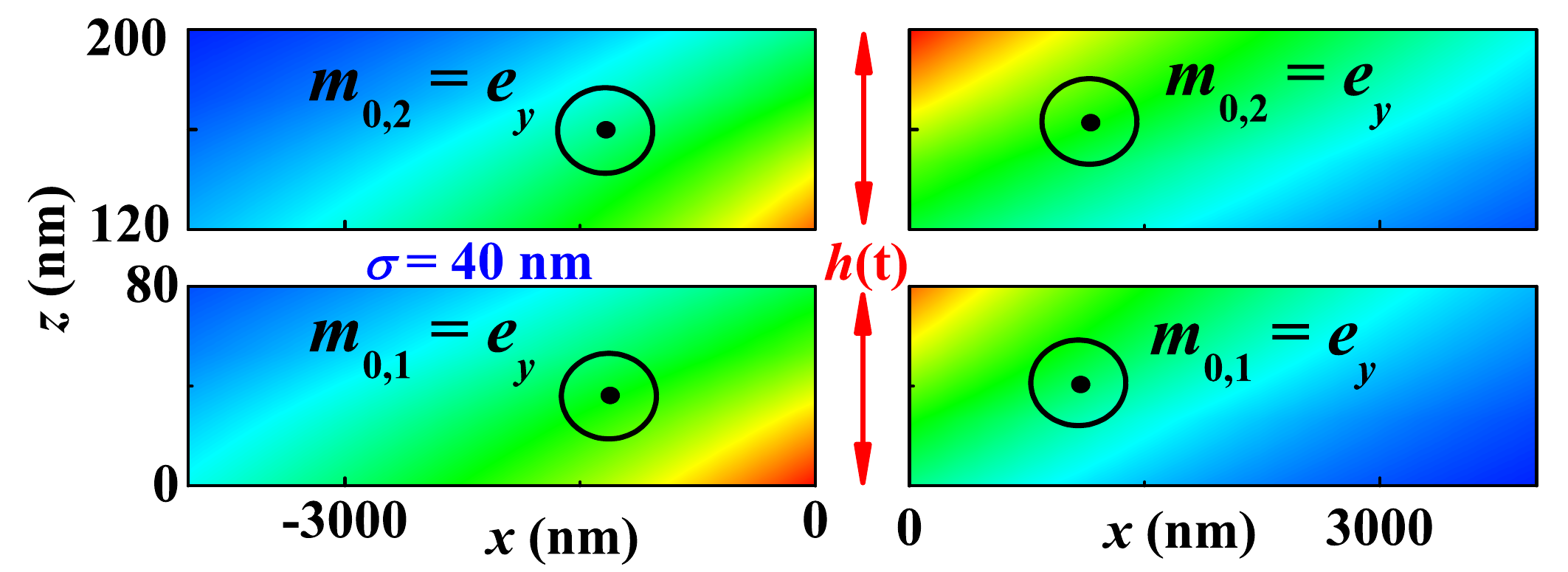}
        \caption{\label{pal-amp} Spatial profiles of MSSWs amplitudes propagating in two films with parallel ground state magnetization without electric field ($ E_z = 0 $). The local microwave field $ h(t) $ is applied at the edges of both films $x=0$. The frequency of microwave field is equal to 5 GHz.}
\end{figure}}

\textbf{\begin{figure}
        \includegraphics[width=0.48\textwidth]{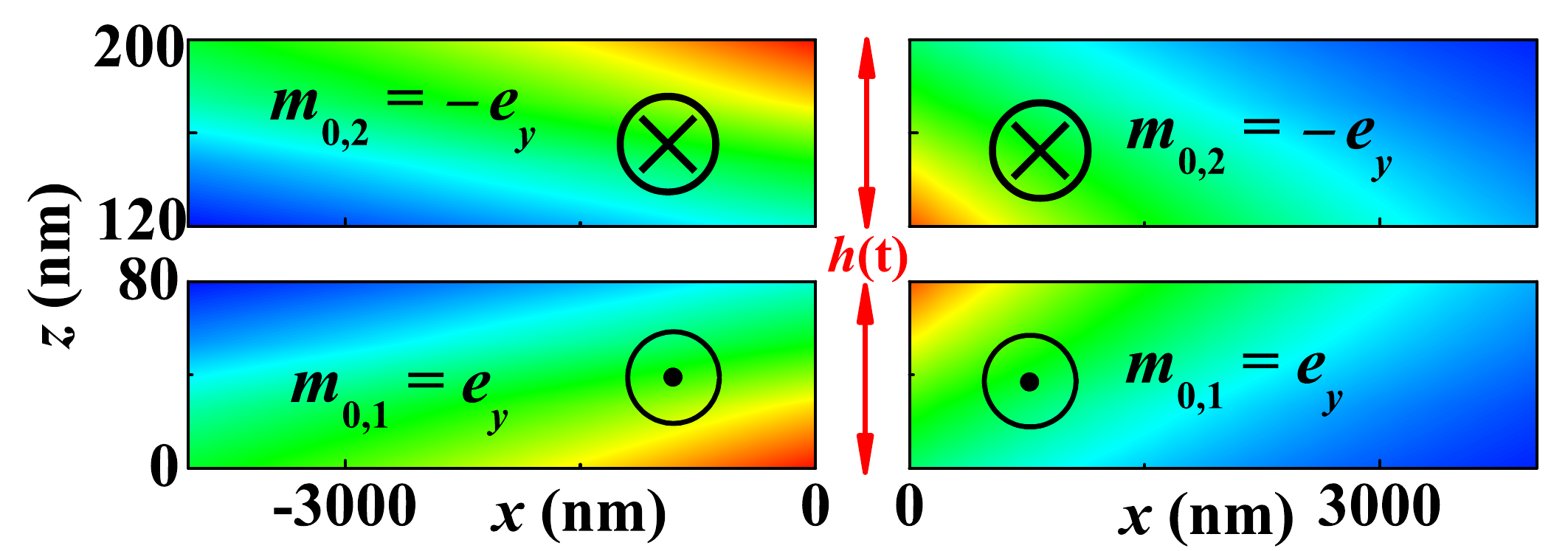}
        \caption{\label{antipal-amp} Spatial profiles of MSSWs amplitudes propagating in two films with antiparallel ground state magnetization. The local microwave field $ h(t) $ is applied at the edges of both films $x=0$. The frequency of microwave field is equal to 5 GHz.}
\end{figure}}

\textbf{\begin{figure}
		\includegraphics[width=0.48\textwidth]{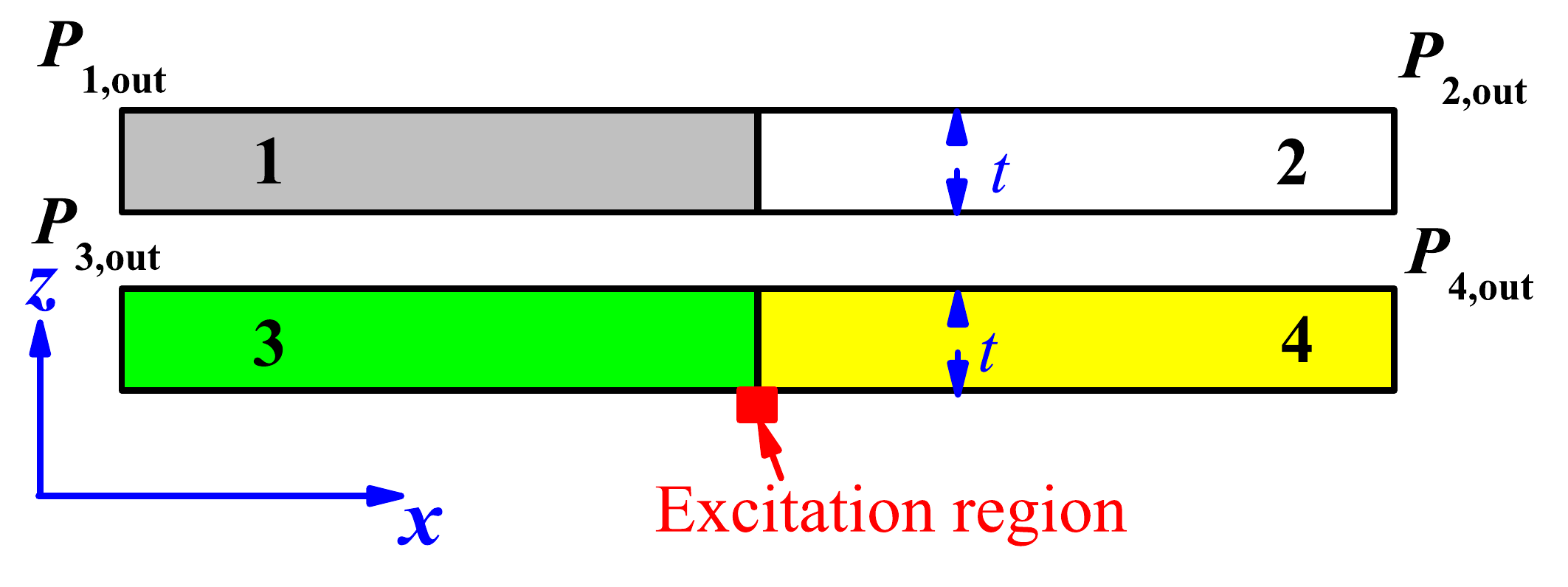}
		\caption{\label{coupler-device}  Schematics of the dipolar coupled SW waveguide. Two magnetic films lying in $ x $-$ y $ plane are separated by a spacer along the $ z $ direction. The SW excitation antenna (red region) is located in the center of the lower magnetic layer. With respect to the center, each magnetic layer is divided into two zones, and the output power of SWs are detected at 4 terminals ($ P_{\mathrm{i,out}} $).}
\end{figure}}

We utilize  Eq. (\ref{LLG}) and numerically explore the dispersion relation of  MSSWs. Simulations are done for a  $ 20 \mathrm{nm} \times 10 \mathrm{\mu m} \times 20 \mathrm{nm} $ unit simulation cell covering  a single film with a size of  $ 30 \mathrm{\mu m} \times 10 \mathrm{\mu m} \times 80 \mathrm{nm} $.
For the numerical integration of the LLG equation (Eq. (\ref{LLG})), we utilize the Dormand-Prince method (RK45) with a fixed time step of 0.5 ps. To calculate the SWs and dispersion curves, at first we let the magnetization to relax to the stationary state. Afterwards the stationary state we consider as the ground state and the SW excitation we add on top of the ground state. To excite SWs in a wide frequency range, we utilize periodic field pulse $ h(t) = h_a \vec{y} \sin(2 \pi f_H t)/(2 \pi f_H t)  $, with the amplitude $ h_a = 1 $ T and cutoff frequency $ f_H = 25 $ GHz, applied locally to the region of the sample $ x = 0 $. We analyze the fluctuation statistics extracted from each cell $ M_x $ during the 200 ns of observation with the time step 20 ps. The frequency resolution of the obtained spectrum is 0.005 GHz. To calculate the SW dispersion relation we use two-dimensional fast Fourier transformation (FFT) $ M_x(k_x,f)=\frac{1}{N}\sum_{i=1}^{N_z}F_2[M_x(x,z_i,t)] $. Here, $ z_i $ is the i-th cell along $ z $ axis, and $ N_z $ is the total cell number along $ z $ axis. Simulations done for the cell size of $ 10 \mathrm{nm} \times 10 \mathrm{\mu m} \times 10 \mathrm{nm} $ (not shown) lead to identical results.
The local equilibrium magnetization is parallel to the $ y $ axis and MSSW propagates along the $ x $ axis.

An applied electric field (along the $z$ axis) causes in an asymmetry in the spin wave (SW) dispersion relations of MSSWs. The results for the single film is shown in Fig. \ref{sim-dispersion}. The positive electric field $ E_z $ shifts the dispersion relation towards the lower left side, while the negative $ E_z $ shifts it towards the lower right side.

The dispersion relation of MSSW, (see Fig.\ref{sim-dispersion}) calculated for the single film has a minimum in the point $ k_x = 0 $. Profiles of amplitudes of propagating MSSWs are presented in Fig. \ref{single-amp}. As we see, the amplitude of the MSSW propagating in the $ +x $ direction has a maximum in the upper surface ($ z = 80 $ nm)  Fig. \ref{single-amp},  whereas in contrast amplitude of the MSSW propagating in the reversed direction $ -x $ has a maximum in the lower surface.  Reversing of the local magnetization switches the upper/lower surfaces (not shown). Thus localization of the MSSW either in the upper/lower surfaces has a chiral character.

As a next step, we consider two coupled MSSW wave guides and analyze the MSSW dispersion relation for two coupled wave guides. A spacer of the thickness $ \sigma = 40  $nm is inserted between the two films of the same thickness $ t = 80 $ nm. The dipolar coupling between the films leads to the formation of the "acoustic" and "optic" modes \cite{prb14950,prb144425}. The dispersion relation of the coupled MSSW modes is shown in Fig. \ref{sim-dispersion-coupled}. When two films have the same equilibrium magnetization (Fig. \ref{sim-dispersion-coupled}(b)),  the dispersion relations have an axial symmetry with respect to the point $ k_x = 0 $. In both films,  the amplitudes of MSSWs are stronger at upper or lower surfaces, depending on the orientation of the equilibrium magnetization and the propagation direction  (Fig. \ref{pal-amp}). For two films with  opposite magnetizations (Fig. \ref{sim-dispersion-coupled}(c)), the dispersion relations become rather different and asymmetric. The gap between the two MSSW modes is much larger for the MSSWs propagating in the negative $ -x $ direction. Due to opposite magnetizations, the stronger MSSWs are located at different surfaces of the two films, as is shown in Fig. \ref{antipal-amp}. Switching of the magnetization direction either in the case of a single or for the two parallel films has no impact on the MSSW dispersion relations, while in the antiparallel films the dispersion is flipped relative to the point  $ k_x = 0 $. Hence, the MSSW dispersion relation becomes chiral and nonreciprocal when the films are antiparallel.

\section{Analytical model}

\begin{figure}
    \includegraphics[width=0.48\textwidth]{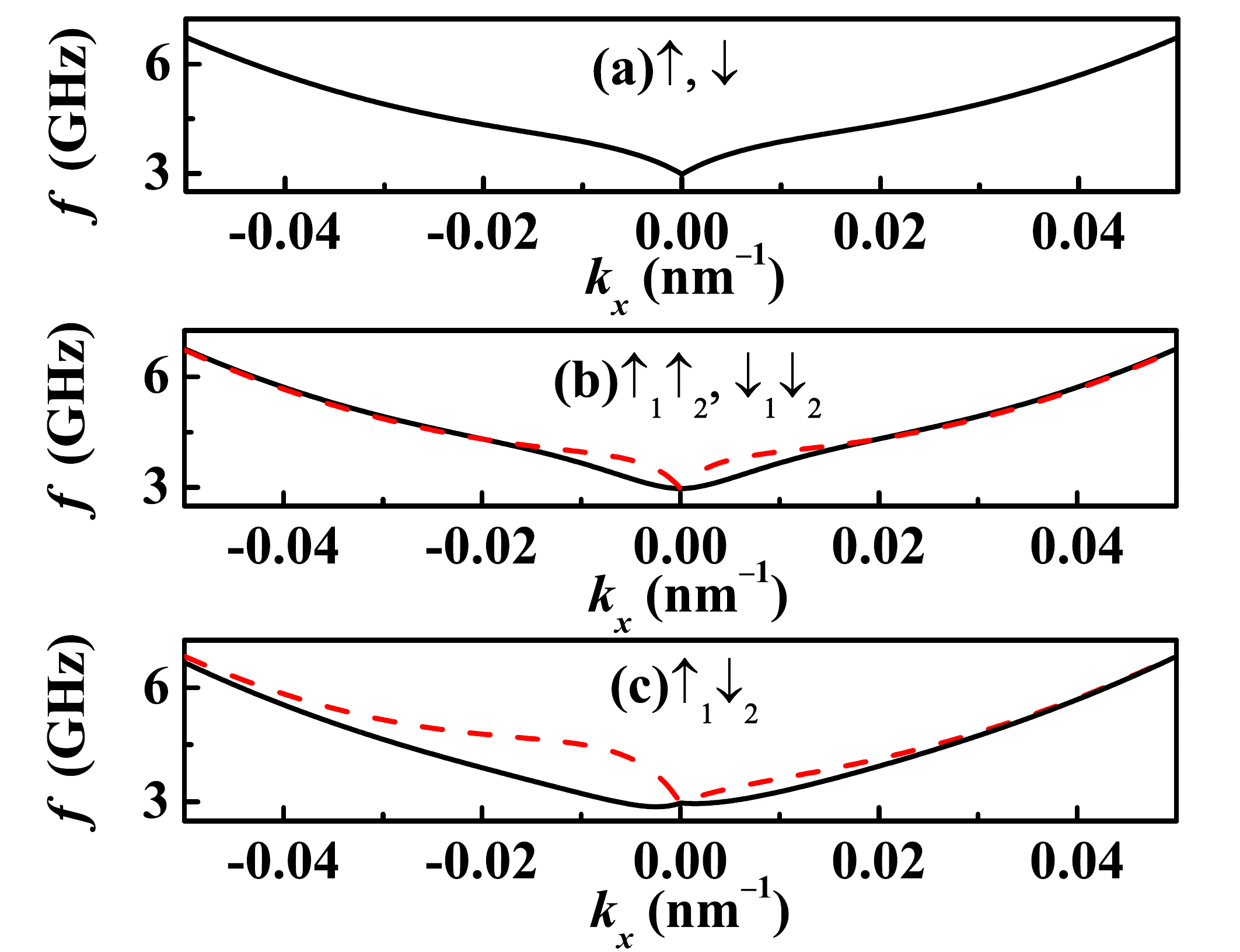}
    \caption{\label{ana-dispersion-coupled} The spin-wave dispersion relation, analytical result: (a) the single film (magnetization is aligned along the $ + \vec{e}_y $ ($ \uparrow $) or $ -\vec{e}_y $ direction ($ \downarrow $)), (b) two parallel coupled films ($ \vec{m}_{0,1} $ and $ \vec{m}_{0,2} $. Magnetization in both films is aligned along the $ \uparrow \uparrow$ or $ \downarrow\downarrow$) directions and (c) two coupled films, directions of magnetization in films are antiparallel ($ \vec{m}_{0,1} = \uparrow $ and $ \vec{m}_{0,2} = \downarrow $).}
\end{figure}

We develop  a simple analytical model to understand the MSSW dispersion relations for both single and coupled magnetic films.  We apply the constant electric field along the $ z $ axis $ \vec{E} = (0, 0, E_z) $, while the static equilibrium magnetization is oriented parallel to the $ y $ axis $ \vec{m}_{0,p} = \pm\vec{e}_y $.
Magnetization vectors in the first and the second $ p = 1, 2 $ films have the form $ \vec{m}_p(r, t) = \vec{m}_{0,p} + \vec{m}_{sw,p} \mathrm{exp}(i(\vec{\kappa} \cdot \vec{r}+\omega t)) $. Here, $ \vec{m}_{sw, p} \ll 1 $ is the small deviation of magnetization from equilibrium. The wave vector $ \kappa $ is a sum of the in-plane wave vector $ \vec{k} = k_x \vec{e}_x + k_y \vec{e}_y $ and the perpendicular (to the \textit{xy} plane) wave vector $ k_z $. We insert the magnetization vector $ \vec{m}_p(r, t)$ into the Eq. (\ref{LLG}) and deduce:

 \begin{equation}
 \displaystyle i(\omega - \omega_{E}) \vec{m}_{k,p} = \vec{m}_{0,p} \times \sum_{q} \hat{\vec{\Omega}}_{pq}\cdot m_{sw,q}.
 \label{FFT}
 \end{equation}
Here, $ p,q=1,2 $ enumerates the layers, and the tensor $ \hat{\vec{\Omega}}_{pq} $ has the form:
\begin{equation}
\displaystyle \hat{\vec{\Omega}}_{pq} = \omega_{ex} \delta_{pq}\hat{\vec{I}} + \omega_{M} \hat{\vec{F}}(d_{pq}).
\label{tensorO}
\end{equation}
Here, for the sake of brevity we introduce the following notations: $ \omega_{ex} = \gamma H_0 + a_{ex} k^2 $, $ a_{ex} = 2 \gamma A_{ex} /(\mu_0) M_s $, $ \omega_{M} = \gamma M_s $, and  $ \omega_{E} = 2 (\vec{m}_{0,p}\cdot \vec{e_y}) \gamma c_E E_z k_x /(\mu_0 M_s)$. The wave vector $ k = \sqrt{k_x^2 + k_y^2} $, the distance between two wave guides $ d_{pq}=t + \sigma $, and $ \sigma $ is the gap between the wave guides. The dynamic magneto-dipolar interaction is described by the tensor\cite{e1701517, Beleggia,Verba2012} $ \hat{\vec{F}} $:
\begin{equation}
\begin{aligned}
\displaystyle \hat{\vec{F}}(d_{pq}) &= \int\hat{\vec{N}}(d_{pq}) e^{i \vec{k} \cdot \vec{r}} \frac{d^2\vec{k}}{(2\pi)^2},\\
              \vec{N}_{\alpha\beta}(d_{pq}) &= \frac{1}{h}\int D_p(k_z)D_q^*(k_z)\frac{\kappa_\alpha \kappa_\beta}{\kappa^2} e^{ik_z d_{pq}} \frac{dk_z}{2\pi}.
\label{tensorF}
\end{aligned}
\end{equation}
Here, the "shape amplitude" $ D_p(k_z)=\int_{-h_s/2}^{h_s/2} m(z) e^{-ik_z z} dz $ describes the influence of the finite thickness of the thin film. The pictorial plot of the system is shown in Fig. \ref{coupler-device}. The SW waveguide consists of two thin magnetic films positioned in the $ x $-$ y $ plane. The thickness of the films is $t$. In  $Z$ direction the films are separated by  a spacer with thickness $\sigma$. The distance between the centers of films is $d_{12}=t+\sigma$. Each film is divided into two regions. The output power of the SW is detected in four ports $P_{i}$. The spin wave is excited by  an antenna attached to the bottom of the lower film (red region). MSSWs mainly are located near the upper or lower surface  (cf. Fig. \ref{single-amp}). We  introduce thus a localization thickness $ h_s $ which is shorter than the real thickness $ t $. Due to the dipolar coupling at the boundaries, the thicknesses of the MSSW profiles in different films are identical and pinned together. Therefore, we use the same ansatz $ m(z) \sim \cos(k_z^p z) $ to describe the induced nonuniform thickness profiles in both films.

\begin{figure}
    \includegraphics[width=0.48\textwidth]{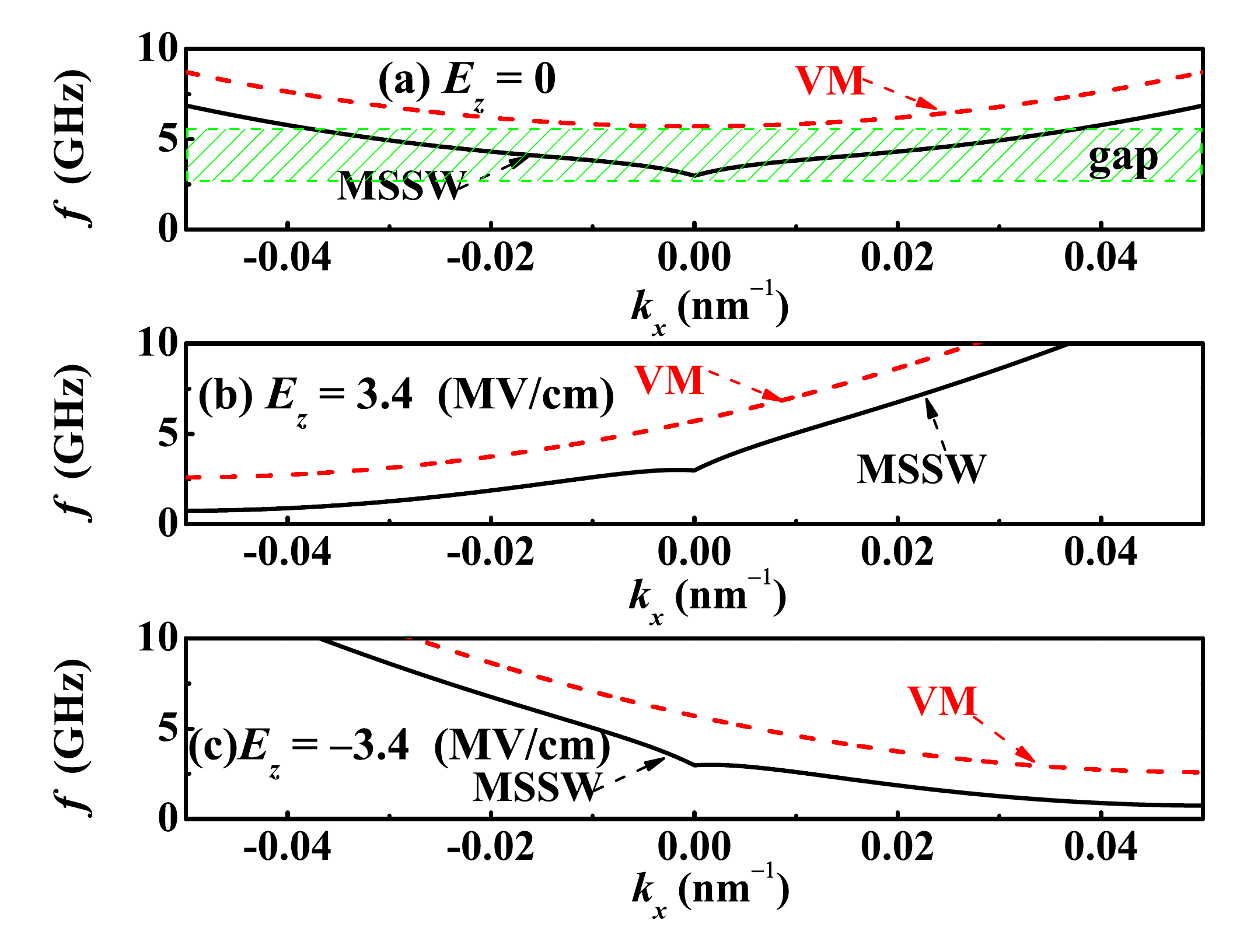}
    \caption{\label{ana-dispersion} Analytically obtained spin-wave dispersion relations for the applied electric field: $ E_z = 0 $ (a), $ E_z = 3.4 $ MV/cm (b) and $ E_z = -3.4 $ MV/cm (c). }
\end{figure}

From Eq. (\ref{FFT}), we obtain the expressions for the dispersion relations of the MSSW modes. The dispersion relation for the MSSW mode in the isolated film reads
\begin{equation}
\begin{aligned}
\displaystyle \omega(\vec{k}) &=  \sqrt{\Omega^{xx}\Omega^{zz}} + \omega_{E}\\
&=  \sqrt{[\omega_{ex} + \omega_{M}F^{xx}(0)][\omega_{ex} + \omega_{M}F^{zz}(0)]} + \omega_{E} .
\label{insolated-dispersion}
\end{aligned}
\end{equation}
From Eq. (\ref{insolated-dispersion}), we evaluate the difference between frequencies of counter propagating spin waves, $ \Delta \omega = \left| \omega(\left|k_x\right|) - \omega(-\left|k_x\right|) \right| = 4 \gamma c_E E_z k_x / (\mu_0 M_s) $. This result is similar to earlier studies \cite{KaiDi}.
The dispersion relation for two coupled layers with the same local magnetization is
\begin{equation}
\displaystyle \omega(\vec{k}) =  \sqrt{[\Omega^{xx} \pm \omega_{M}F^{xx}(d_{12})][\Omega^{zz}\pm \omega_{M}F^{zz}(d_{12})]} + \omega_{E}.
\label{coupled1}
\end{equation}
If the magnetizations of the layers are opposite to each other, the dispersion relation takes the form
 \begin{equation}
\begin{aligned}
\displaystyle &\omega(\vec{k}) = \large[ \omega_{E}^2 + \Omega^{xx}\Omega^{zz}-F^{xx}(d_{12})F^{zz}(d_{12})\omega_{M}^2\pm \\
&(4\omega_{E}^2\Omega^{xx}\Omega^{zz} + \omega_{M}^2 (F^{zz}(d_{12})\Omega^{xx}-F^{xx}(d_{12})\Omega^{zz})^2)^{\frac{1}{2}}\large]^{\frac{1}{2}}.
\label{coupled2}
\end{aligned}
\end{equation}
With these analytical expressions we insert the same material parameters, as used for numerics, and calculate the MSSW dispersion relations. For the profile thickness of the fundamental surface wave, we take $ k_z^p = 0 $.
To simplify the calculation, for the single film and for parallel films we use $ h_s = t $.  The analytical results for the MSSW dispersion relations plotted in  Fig. \ref{ana-dispersion-coupled} is in a good agreement with numerical results.  In the case of the films with antiparallel ground state magnetization (see Fig. \ref{antipal-amp}), the thickness of the localized MSSW depends on the propagation direction, and the localization thickness$ h_s $  is smaller when the MSSW propagates in the $ +x $ direction. Thus, using different $ h_s $ for different propagation directions, we obtain distinct MSSW dispersions, as shown in Fig. \ref{ana-dispersion-coupled}(c). Reversing of the magnetization in both films reverses the localization thickness $  h_s $  of the MSSW propagating in the $ +x $ direction, and flips the MSSW dispersion relations.  Despite the relative simplicity and approximations, our analytical model and obtained results are in a good agreement with the exact results of the micromagnetic numeric simulation. In both cases, we observe that an applied electric field leads to the asymmetric MSSW  dispersion relations, see Figs. \ref{sim-dispersion} and \ref{ana-dispersion}. The slight quantitative difference between the numeric and analytical results arises due to the approximations considered in the analytic part.

\section{Spin wave propagation in the waveguide with defect}

\begin{figure}
    \includegraphics[width=0.48\textwidth]{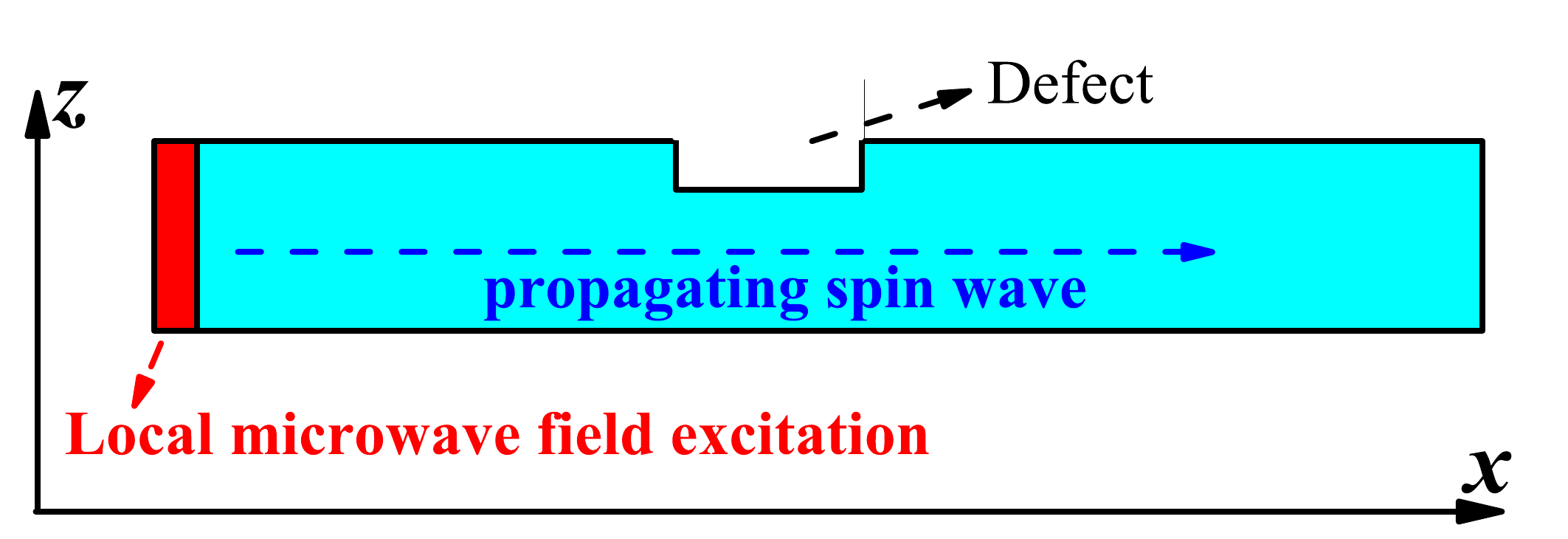}
    \caption{\label{model} Pictorial representation  of  MSSW  waveguide with thickness $ t $. The waveguide film is aligned in the $ x $-$ y $ plane. A surface defect is embedded in the center of the waveguide. The spin-wave propagates along the $ +x $ axis and is excited locally at the left edge of the waveguide. }
\end{figure}

\begin{figure}
    \includegraphics[width=0.48\textwidth]{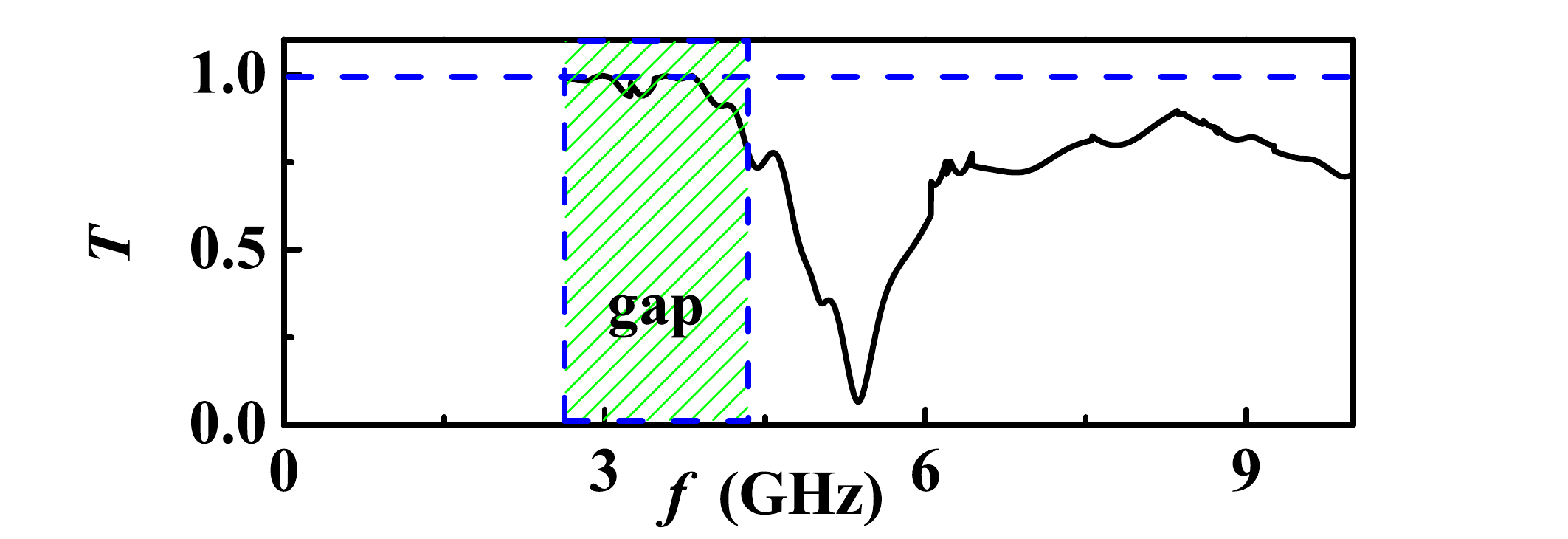}
    \caption{\label{trans-e0} The transmission ratio $ T $ of the MSSW as a function of frequency. The MSSW propagates through the region with the defect. No electric field is applied  ($ E_z = 0 $). The value  $T = 1$ corresponds to the perfect transmission.}
\end{figure}

\begin{figure}
    \includegraphics[width=0.48\textwidth]{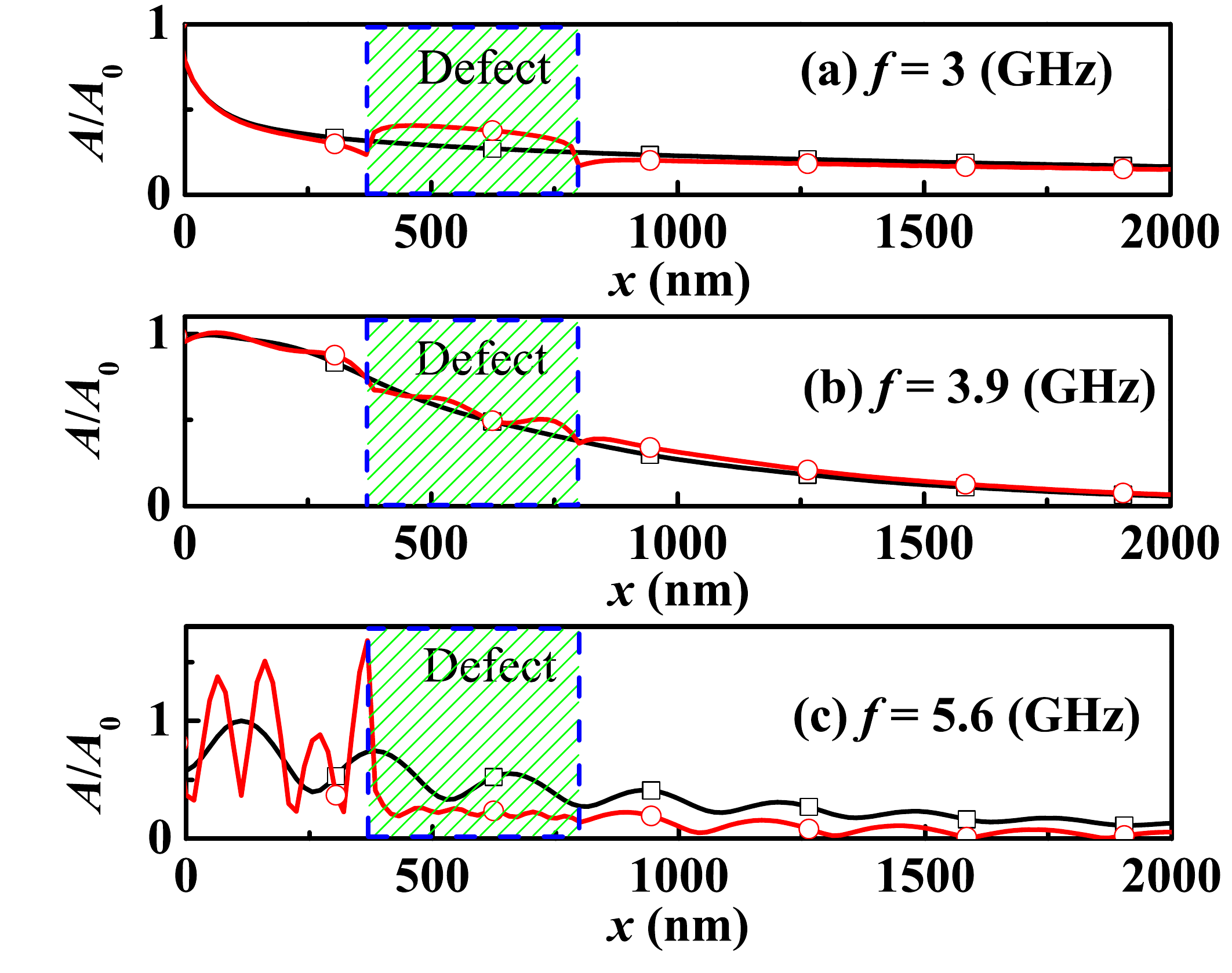}
    \caption{\label{x-e0} Spatial distributions of normalized amplitudes $ A(x) / A_0 $ of excited spin-waves along $ x $. Waves are located on the surface of the waveguide. Frequencies of the waves are $ f = 3 $ GHz (a), 3.9 GHz (b) and 5.6 GHz (c). The amplitude $ A $ is normalized to the maximum value $ A_0 $ of black squares line (when there is no defect), and red circles line represents the MSSW amplitude with defect. The electric field is zero $ E_z = 0 $ .}
\end{figure}

\begin{figure}
    \includegraphics[width=0.48\textwidth]{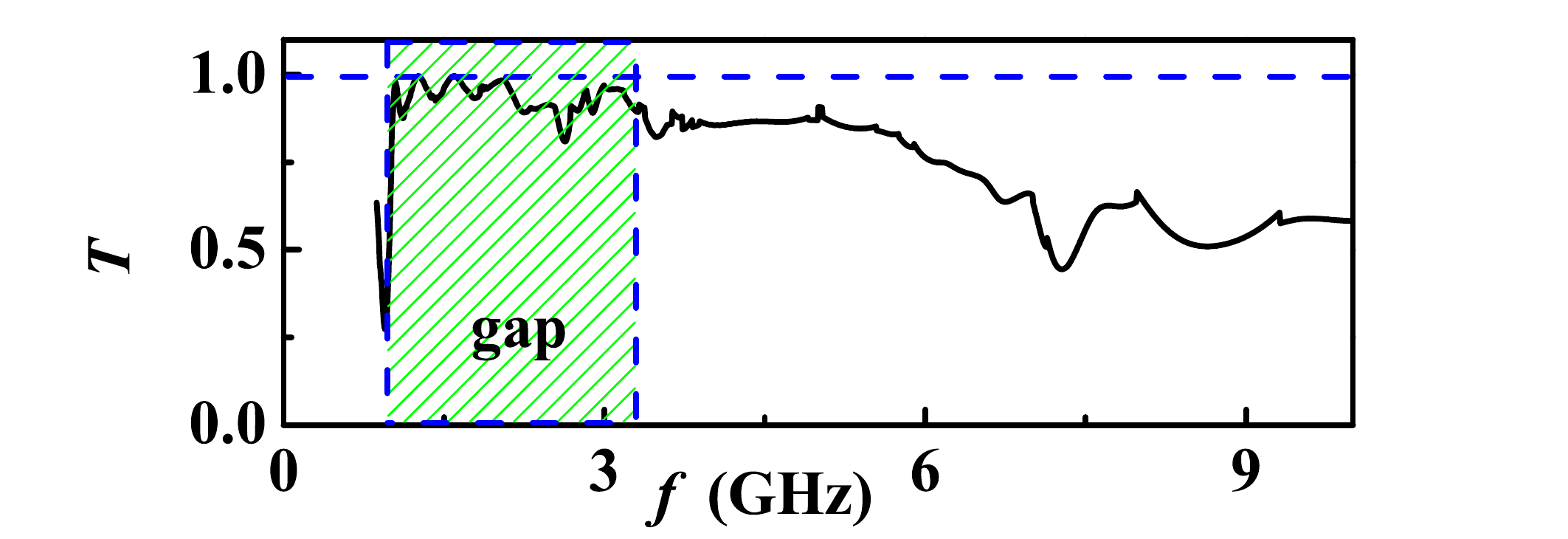}
    \caption{\label{trans-e+}
    The transmission ratio $ T $ of the MSSW as a function of frequency. The MSSW propagates through the region with defect and electric field is zero $ E_z = 3.4$ MV/cm.}
\end{figure}

\begin{figure}
    \includegraphics[width=0.48\textwidth]{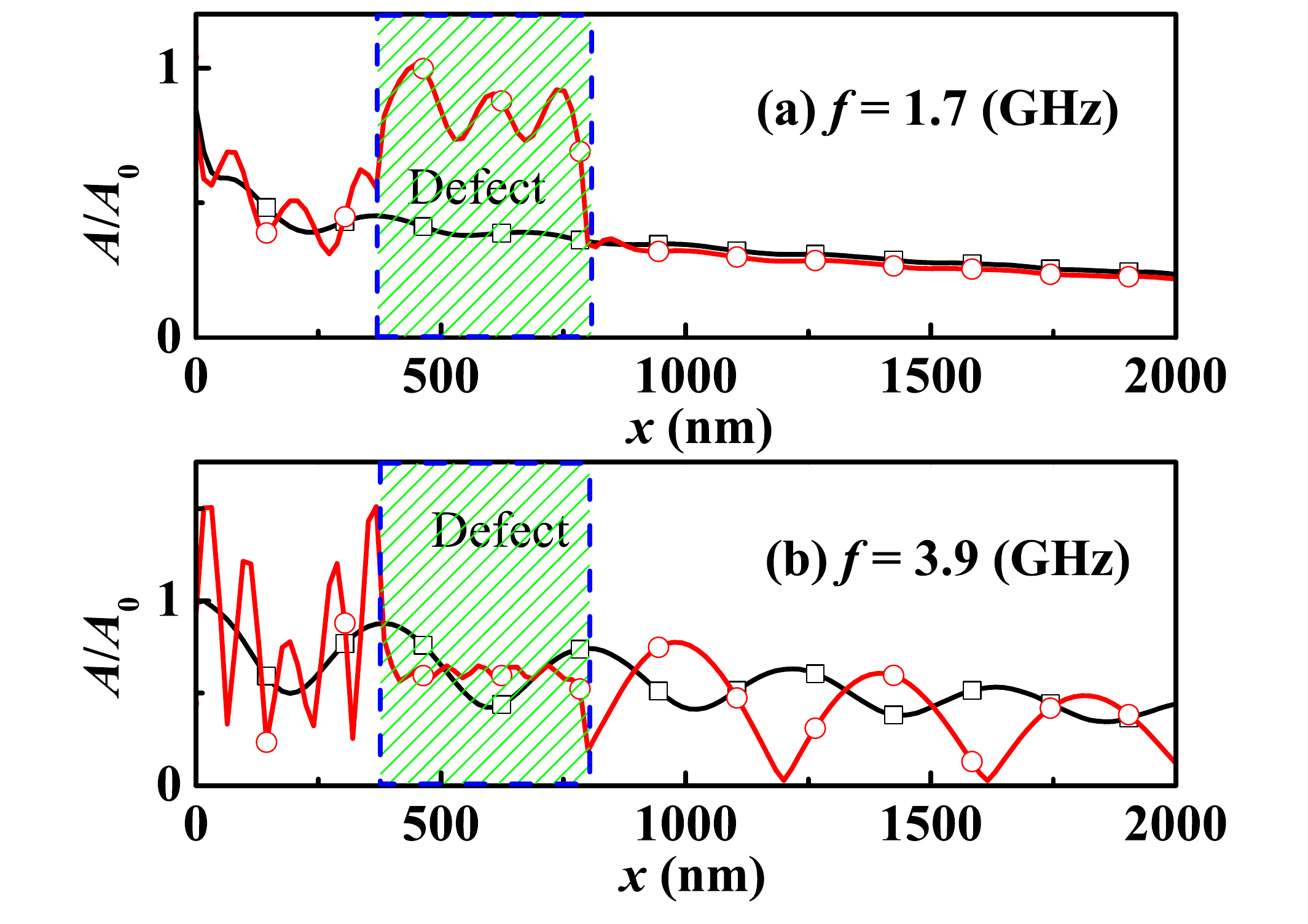}
    \caption{\label{x-e+} Spatial distributions of the normalized amplitudes $ A(x) $ of spin waves. Waves are located  on the surface of the waveguide.
Frequencies of the waves (a) $ f = 1.7 $ GHz  and (b) 3.9 GHz. The applied electric field $ E_z = 3.4 $ MV/cm.}
\end{figure}

The chiral MSSWs are protected against surface inhomogeneities and defects if the frequency of MSSW matches the band gap \cite{Mohseni}. The width of the band gap is quantified in terms of the difference between minimum threshold frequencies of VM and MSSW mods, as demonstrated in Fig. \ref{ana-dispersion}(a). Our numerical simulations confirm this same result. In the calculation we used the sinc field $ h_a \sin(2 \pi f_H t) / (2 \pi f_H t) \vec{y} $, $ h_a = 1 $ T, and $ f_H = 25 $ GHz applied locally in the left edge of the waveguide (Fig. \ref{model}). Thus, spin waves in the frequency interval of $0<f_H< 25$ GHz are excited. Spin waves with the frequencies lower than the smallest frequency of MSSW mode cannot propagate.

Without electric field ($ \vec{E} = 0 $), the lower frequency bound of MSSW is about 2.8 GHz see Fig. \ref{sim-dispersion}. An embedded defect at the surface of the film induces scattering of the spin wave. The scattering process sharply  depends on the frequency of the spin wave $ f = \omega/(2\pi) $. Comparing the amplitudes of the spin waves in the waveguide with ($ A_{\mathrm{with}} $) and without ($ A_{\mathrm{without}} $) the  embedded defect, we numerically evaluate the transmission ratio $ T $ and quantify the transmission coefficient of spin waves to describe the scattering process (Fig. \ref{trans-e0}). In the gap, $ T $ is about 1, indicating that the bulk of MSSWs can transmit through the defect without reflection, i.e backscattering is not permitted. The transmission $ T $ becomes much smaller than 1 outside the gap, due to the strong reflection from the defect. We also compare the spatial distributions of $ A_{\mathrm{with}} $ and $ A_{\mathrm{without}} $ in Fig. \ref{x-e0}. In the absence of the embedded defect,  the amplitude $ A $ decays in the  $ x $ direction. In addition, at $ f = 5.6 $ GHz (beyond the gap), due to the interaction between two SWs with different  wave-vectors (MSSW and VM), the amplitude $ A $ is not uniform and oscillates with $ x $. After embedding the defect, we see a clear signature  of the reflection process and the scattering from  defects for $ f = 5.6 $ GHz ($ T = 0.34 $), whereas for $ f = 3 $ GHz ($ T = 1 $), the spin wave entirely transmits through the defect. Our results are in a good agreement with the Ref. [\mycite{Mohseni}].

Applying an electric field shifts the gap down and thus decreases the frequency interval with $ T \approx 1 $, as shown in Fig. \ref{trans-e+}. In the frequency interval from $1<f<3$ GHz, the frequency of the MSSW matches the band gap, and therefore MSSW can transmit through the defect entirely without reflection (as demonstrated by the amplitude $ A(x) $ for the frequency $ f = 1.7 GHz $ ($ T = 0.96 $) in Fig. \ref{x-e+}(a)). Outside the band gap, the reflection of the spin wave at the defect becomes significant (Fig. \ref{trans-e+}). The reflection, $ A(x) $ at $ f = 3.9 $ GHz ($ T = 0.85 $) is shown in Fig. \ref{x-e+}(b).

\section{SW transfer between two coupled films: The role of the electric field}

\begin{figure}
	\includegraphics[width=0.48\textwidth]{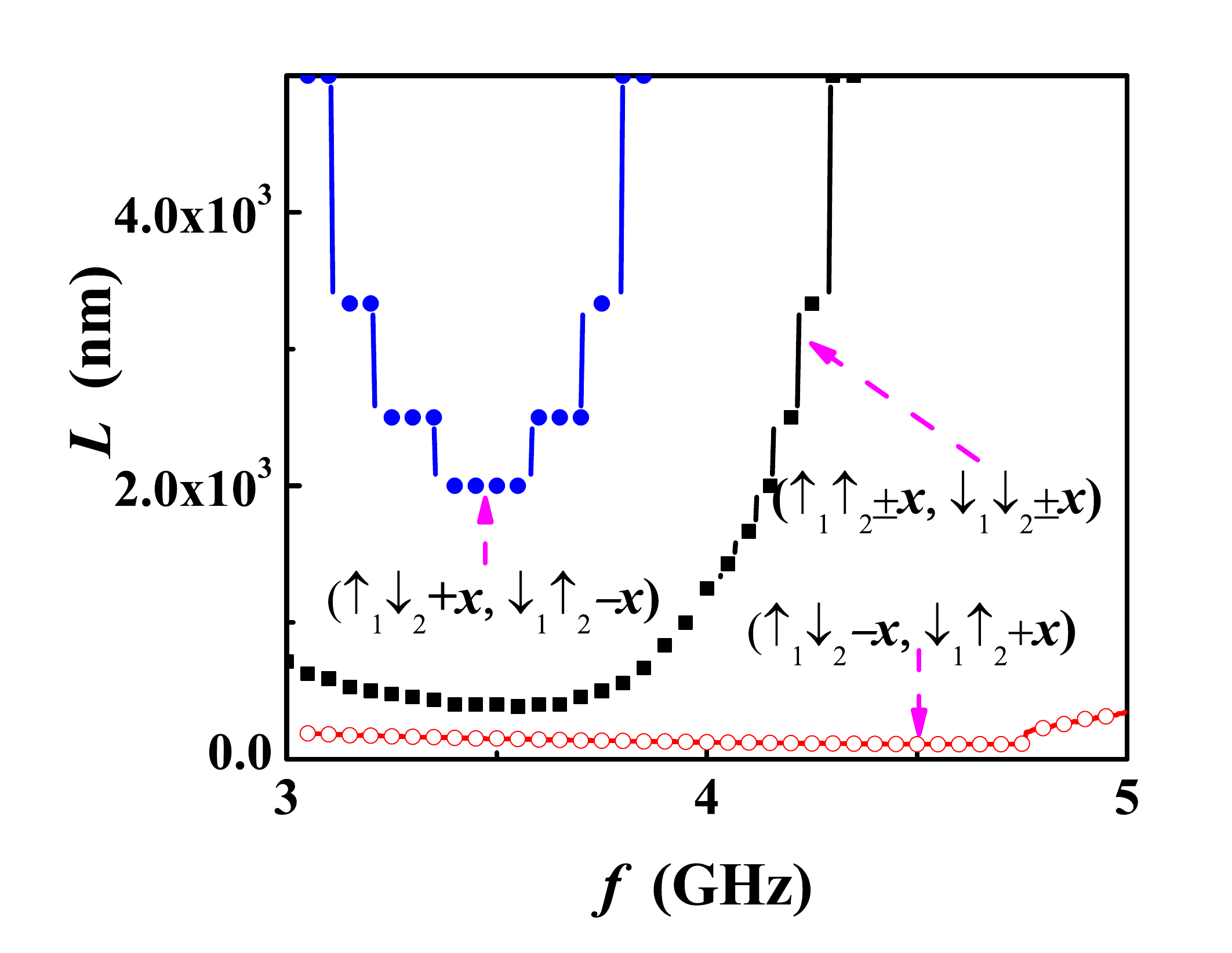}
	\caption{\label{L-period} The coupling length $ L $ between spin waves as a function of the spin wave frequency $ f $. For parallel films $ \uparrow_1 \uparrow_2 $ or $ \downarrow_1 \downarrow_2 $, $ L $ is insensitive  to the propagation direction $ \pm x $  of SWs. In case of anti-parallel magnetizations of the films  $ \uparrow_1 \downarrow_2 $ or $ \downarrow_1 \uparrow_2 $,  the SW's propagation length $ L $ is different for ($ +x $ or $ -x $).}
\end{figure}

\begin{figure}
	\includegraphics[width=0.48\textwidth]{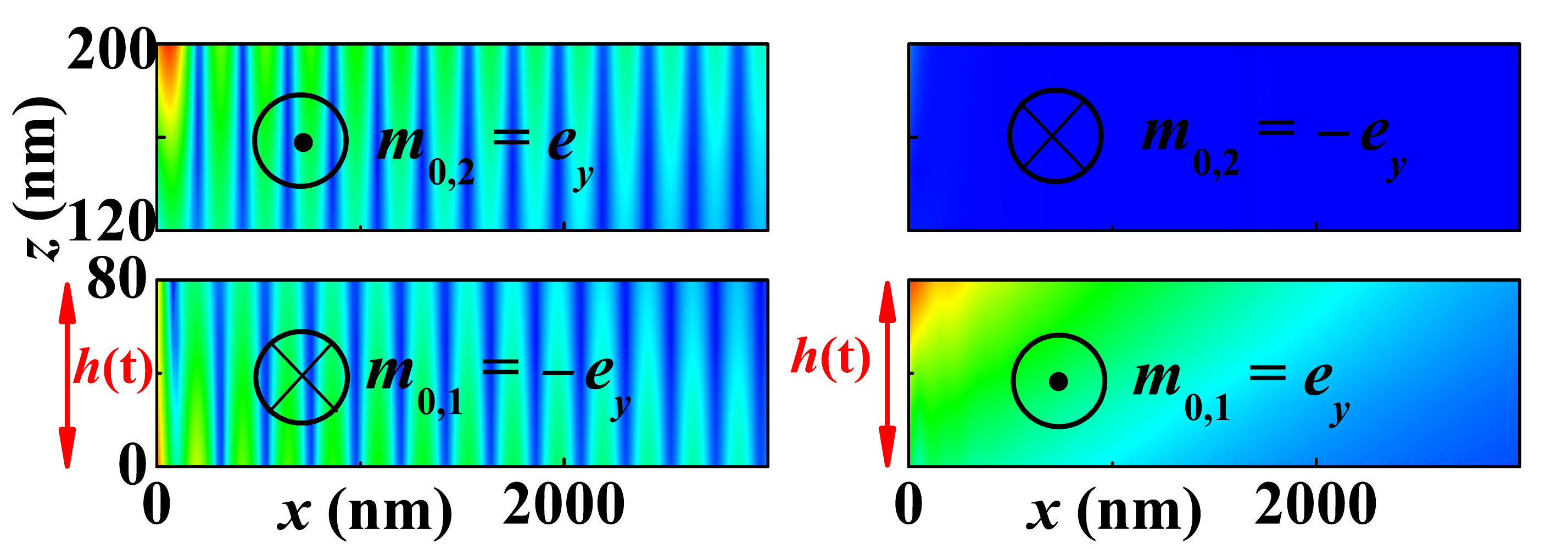}
	\caption{\label{transfer} At 4.5 GHz, spatial profiles of amplitudes of propagating SWs in two antiparallel films. The local microwave field $ h(t) $ is applied in the lower film at $ x = 0 $. Left and right panels correspond to opposite magnetization.}
\end{figure}

We consider the dynamics of simultaneously excited two propagating SW modes. The wave guides are coupled through the dipole-dipole interaction. The dipole-dipole coupling splits the waveguide modes into the symmetric (acoustic) and asymmetric (optical) mods. The acoustic and optical modes may interact and  exchange  energy. The transfer of the energy  takes place on the characteristic length scale $ L = \pi/\Delta k_x $ termed as "coupling length" \cite{e1701517}. The coupling length $ L $ characterizes the propagation length of the SW excitation in the first waveguide, before transferring the energy to the second waveguide, i.e. "mean free path" of the SW. The $ \Delta k_x $ is the difference between the wave vectors of the acoustic and the optical  modes at the same frequency.  The simulated value of the coupling length  $ L$ depends on the frequency $ f $ of the propagating wave, see Fig. \ref{L-period}.  Our results show that the coupling length  $ L$ and the energy transfer process can be controlled through two factors: the direction of the equilibrium magnetization and the direction of propagation of SWs.  For example, for $ f = 4.5 $ GHz, $ L \approx \infty  $ for ($ m_{0,1} = e_y $ and $ m_{0,2} = -e_y $) is much larger than $ L = 100 $ nm for  ($ m_{0,1} =- e_y $ and $ m_{0,2} = e_y $). The energy transfer  can be switched on and off via switching the magnetization directions. The propagating SW excited in the lower film (Fig. \ref{transfer}), is effectively decoupled from the upper film $ L \approx \infty $. The energy transmission is switched as $ L = 100 $ nm after reversing the magnetization.

\begin{figure}
    \includegraphics[width=0.48\textwidth]{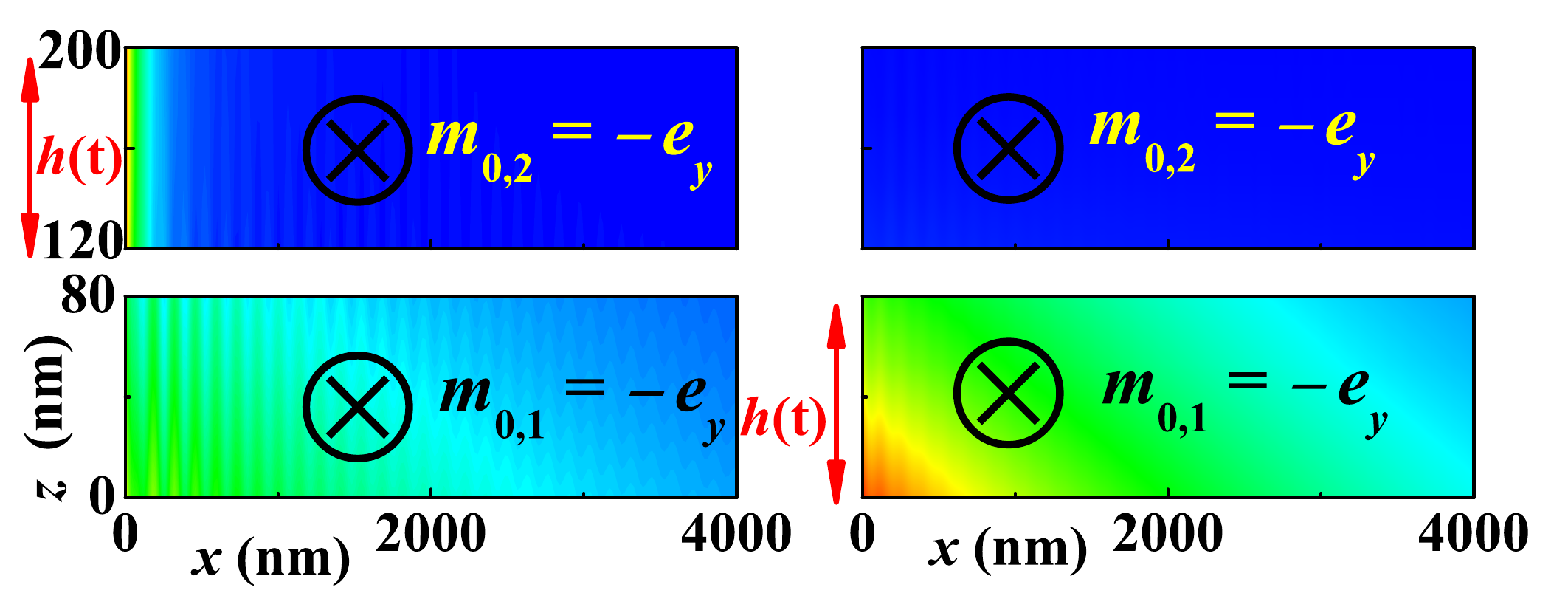}
    \caption{\label{coupler-electric field} The electric field $ E_z = 3.4 $ MV/cm is applied to the bottom layer. The frequency of the local microwave applied to the bottom (or top) film at $ x = 0 $ for the right (or left) panel is $ f = 1 $ GHz. The spatial profiles of propagating MSSW amplitudes are plotted.}
\end{figure}

Besides, an electric field shifts the SW dispersion relation in the coupled films, and this fact is important for controlling the transmission properties of the SWs.  The electric field $ \vec{E} = (0,0,E_z) $ applied only to a single film  shifts selectively the  dispersion relation solely in this layer (similar to Fig. \ref{sim-dispersion}). Therefore, the SW for a frequency lower than 3 GHz can propagate in the selected film, while propagation in the second layer is forbidden. To illustrate this statement, we apply the microwave field of the frequency $ f = 1 $ GHz, to the bottom or top layer in the vicinity of the region $ x = 0 $ and excite the SW.  The constant electric field we apply to the bottom layer only.  SW  propagates in the bottom layer,  while propagation in the top layer is restricted, see Fig. \ref{coupler-electric field}.  Typically, the chirality leads to a breaking of the inversion symmetry.  The dynamically imposed chirality shows the same features.
The effect we observed for SWs is somewhat similar to the phenomenon observed in the context of non-hermetian optics \cite{nat192} (in fact, for narrow waveguides in the paraxial approximation our models can be mapped to the PT-symmetric model used for the coupled photonic waveguides \cite{nat192}), and allows us to switch on/off the propagation of MSSWs in the selected layer.

\section{The designing of the spin wave directional coupler}

The functionality of the reconfigurable nanoscale spin wave directional coupler depends on the structure of the band gap (efficiently controlled through the electric field) and the direction of the ground state equilibrium magnetization. The direction of the magnetization can be switched via the external magnetic field or the current-induced spin-orbit torque (SOT). The SOT has  in addition certain advantages, see Ref. [\mycite{Scientific Reports}] for more details.

\begin{figure}
    \includegraphics[width=0.48\textwidth]{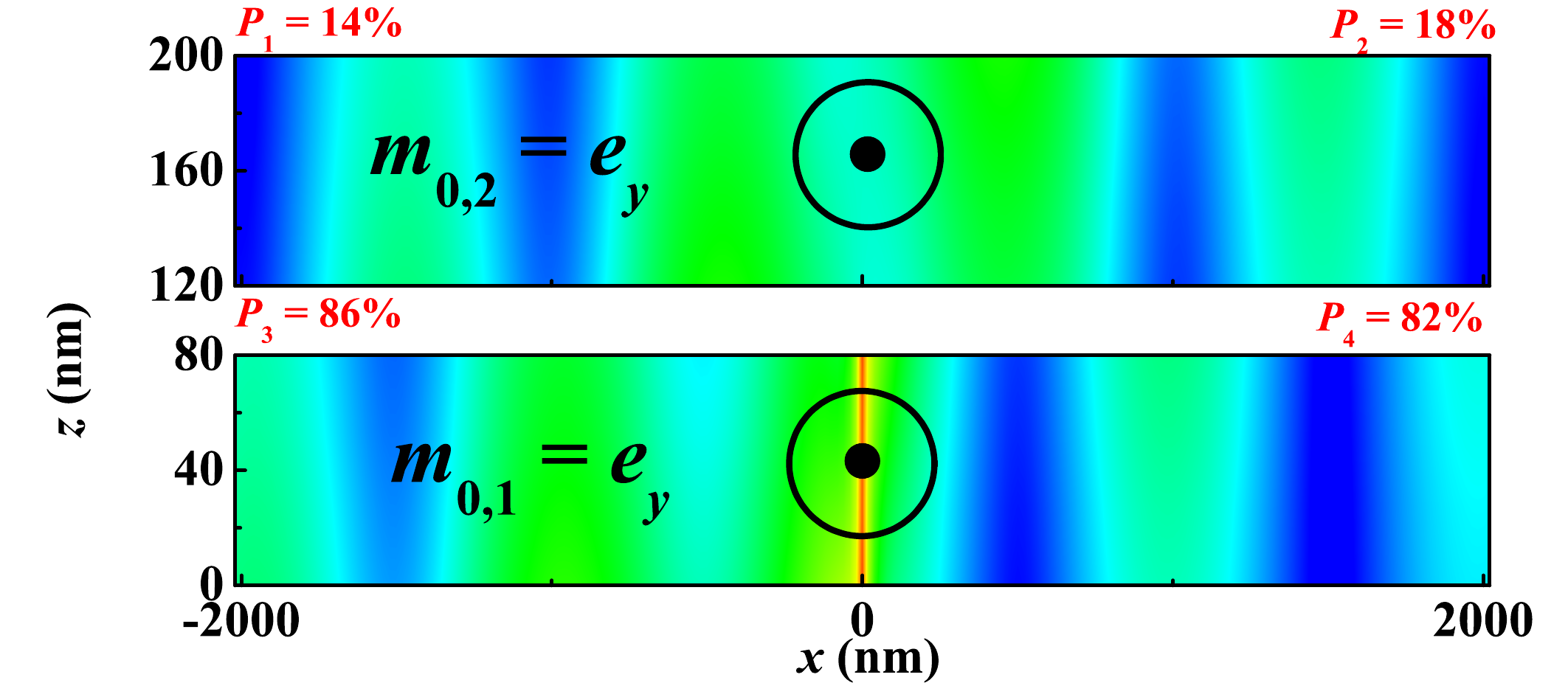}
    \caption{\label{waveguide-pal-641} Spatial profiles of amplitudes of propagating SWs in two parallel films. $ P_1 $ and $ P_3 $ are detected at the left side of the waveguide ($ x = -2000 $ nm), and $ P_2 $ and $ P_4 $ are located at the right side of the waveguide. SWs are excited by the rf field with the frequency  3.2 GHz.}
\end{figure}

\begin{figure}
    \includegraphics[width=0.48\textwidth]{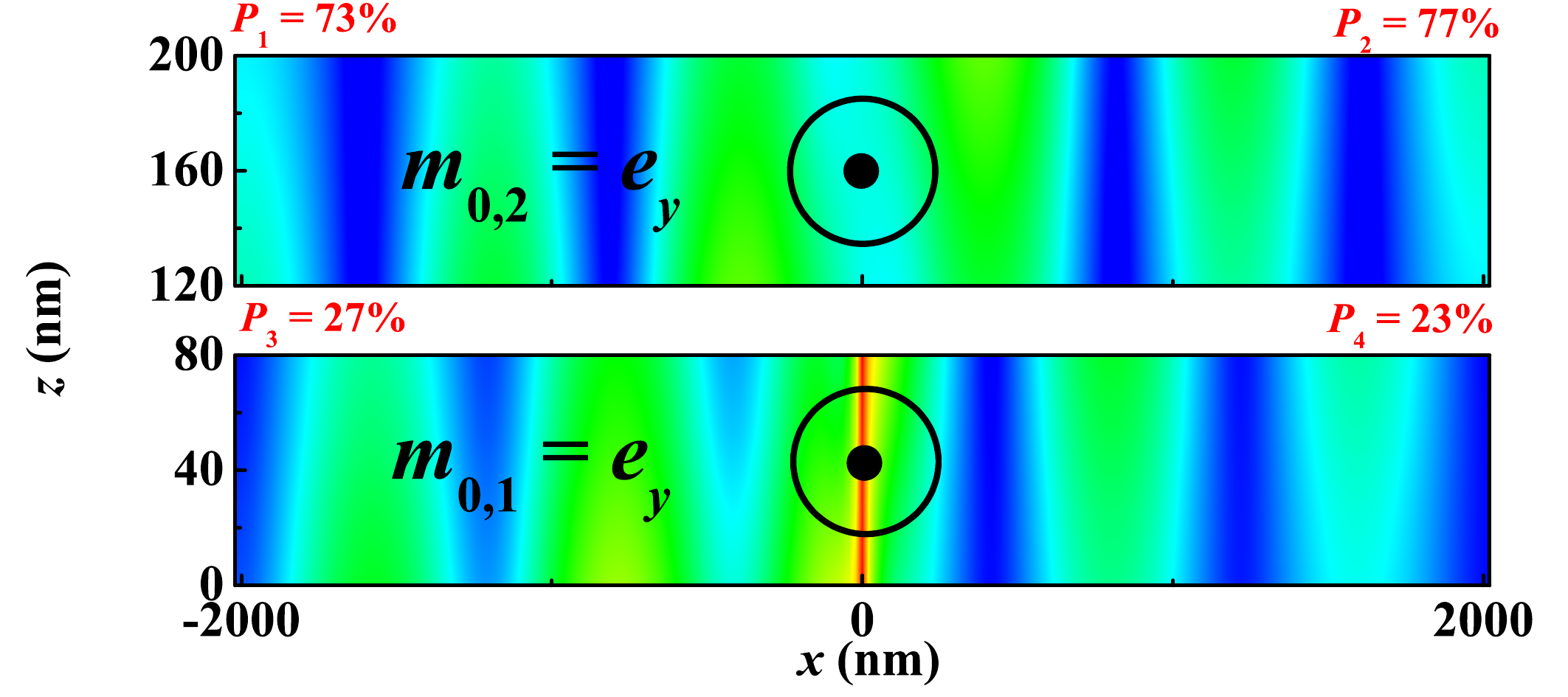}
    \caption{\label{waveguide-pal-689} Spatial profiles of amplitudes of propagating SWs in two parallel films. SWs are excited by the rf field with the frequency  3.44 GHz.}
\end{figure}

\begin{figure}
    \includegraphics[width=0.48\textwidth]{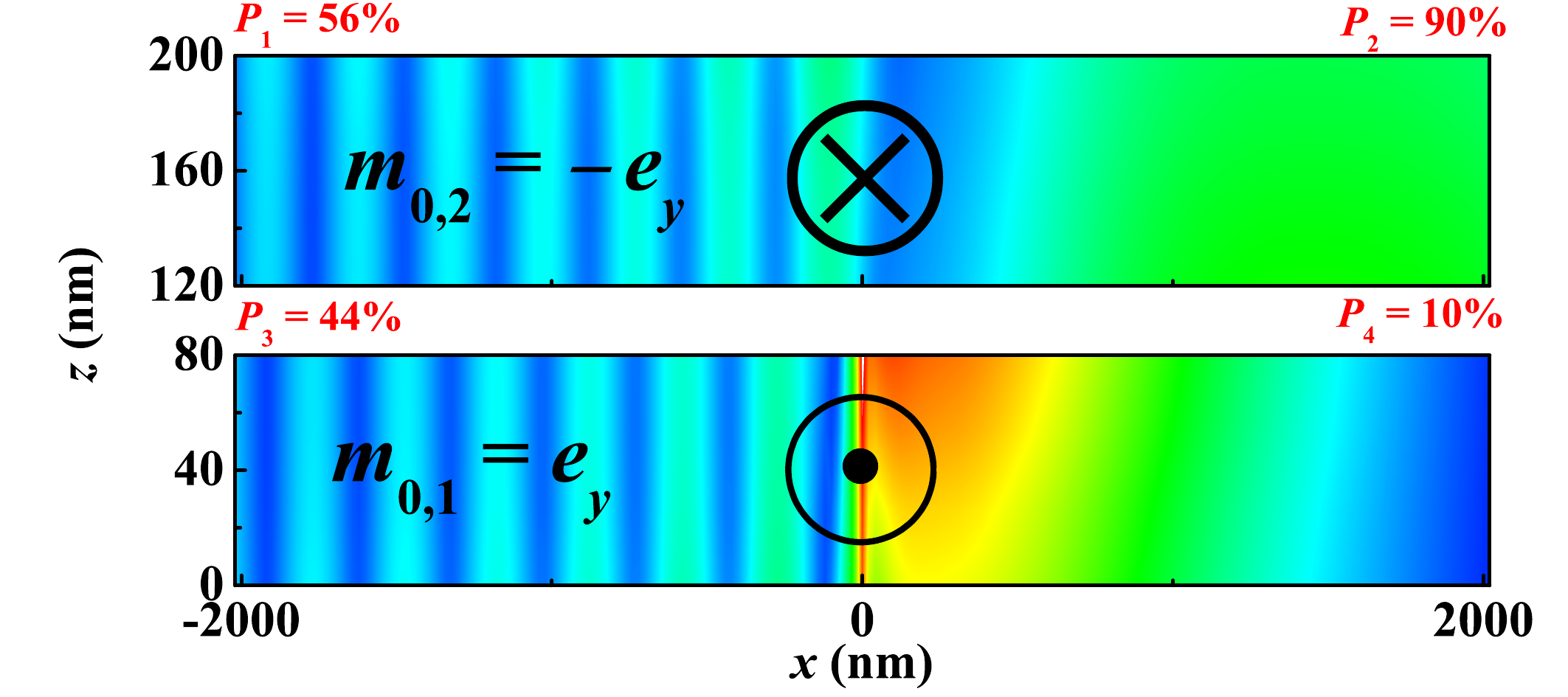}
    \caption{\label{waveguide-anti-707} Spatial profiles of amplitudes of propagating SWs in two antiparallel films. SWs are excited by the rf field with the frequency 3.53GHz.}
\end{figure}

\begin{figure}
    \includegraphics[width=0.48\textwidth]{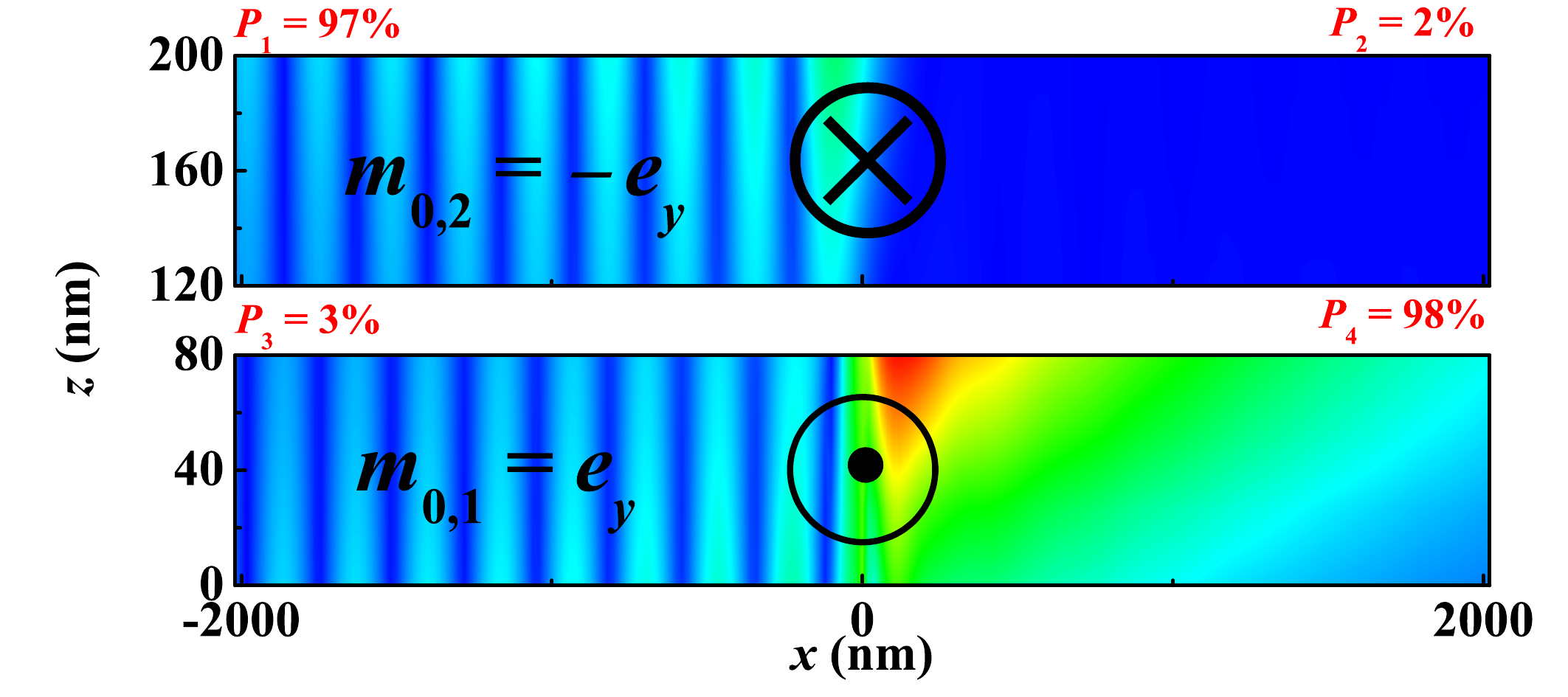}
    \caption{\label{waveguide-anti-848} Spatial profiles of amplitudes of propagating SWs in two antiparallel films. SWs are excited by the rf field with the frequency 4.24GHz.}
\end{figure}

Here, we propose a  reconfigurable directional coupler utilizing SOT for controlling the direction of magnetization:  In parallel coupled films, the dispersion relation, as well as coupling length $ L $, are the same for the opposite propagation directions  of MSSWs. Thus $ P_1 \approx P_3 $ and $ P_2 \approx P_4 $ reach their maxima or minima  synchronous at the same moment of time, as demonstrated in Figs. \ref{waveguide-pal-641} and \ref{waveguide-pal-689}. After switching the magnetization distribution to the anti-parallel direction, the dispersion relation and $ L $ becomes asymmetric (Fig. \ref{L-period}), and the large difference between left side output $ P_1 / P_3 $ and right side output $ P_2 / P_4 $ develops, as is demonstrated in Figs. \ref{waveguide-anti-707} and \ref{waveguide-anti-848}.

\begin{figure}
    \includegraphics[width=0.48\textwidth]{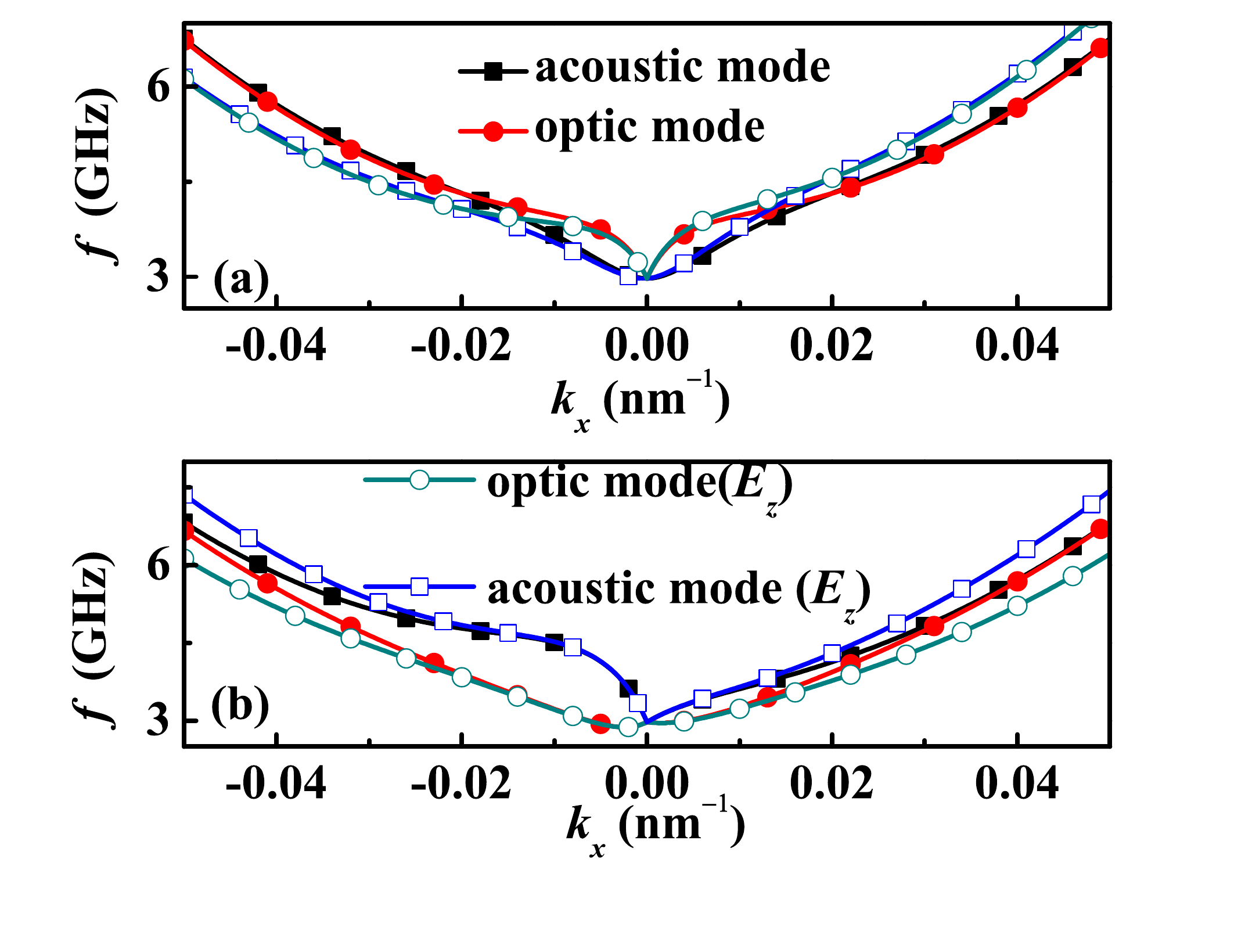}
    \caption{\label{ana-ele-dispersion} Analytically derived  spin-wave dispersion relations with $ E_z = 0.34 $ MV/cm (open dots) and without electric field (solid dots), in the coupled parallel (a) and antiparallel (b) films. }
\end{figure}

\begin{figure}
    \includegraphics[width=0.48\textwidth]{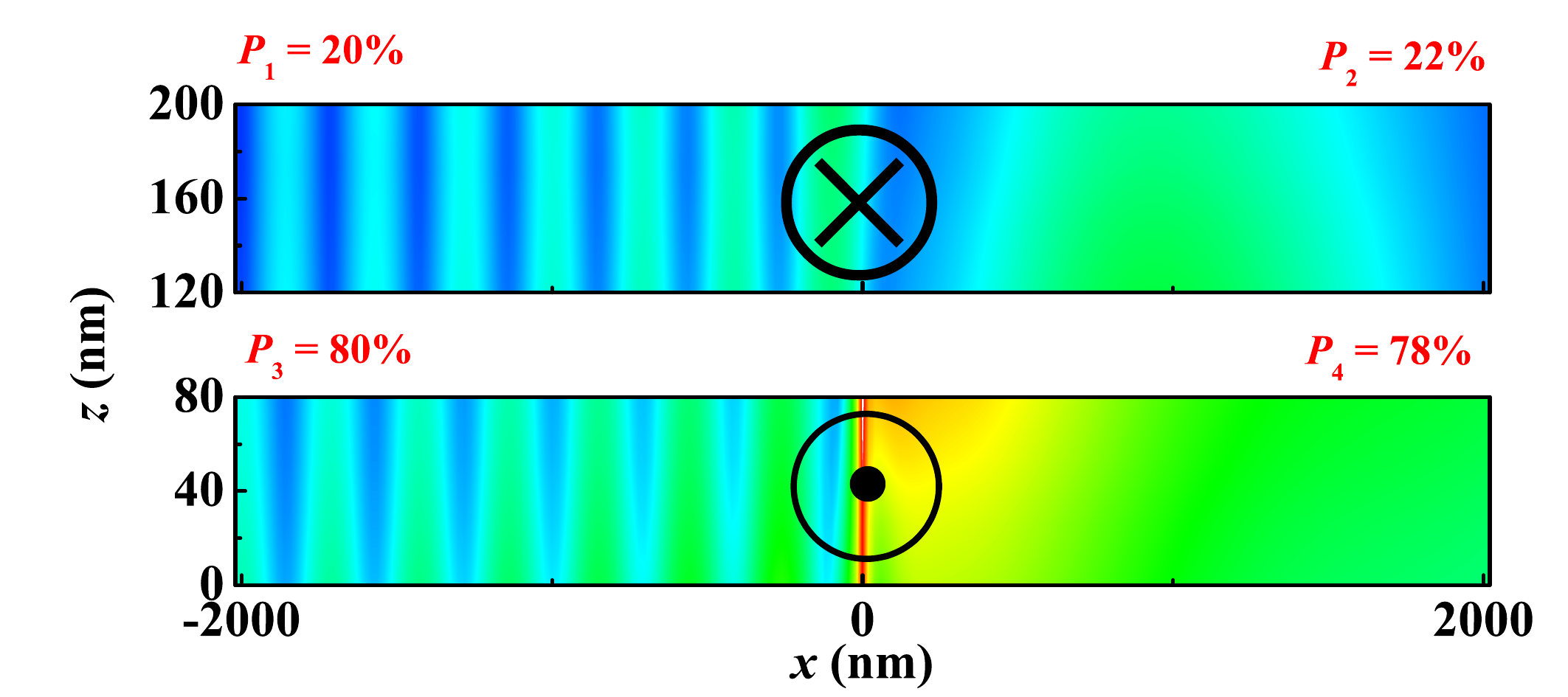}
    \caption{\label{waveguide-anti+ey-707} Spatial profiles of amplitudes of propagating SWs in two antiparallel films. The frequency of the rf field is 3.53 GHz, and the amplitude of the external electric field $ E_z = 0.34 $ MV/cm.} \end{figure}

\begin{figure}
    \includegraphics[width=0.48\textwidth]{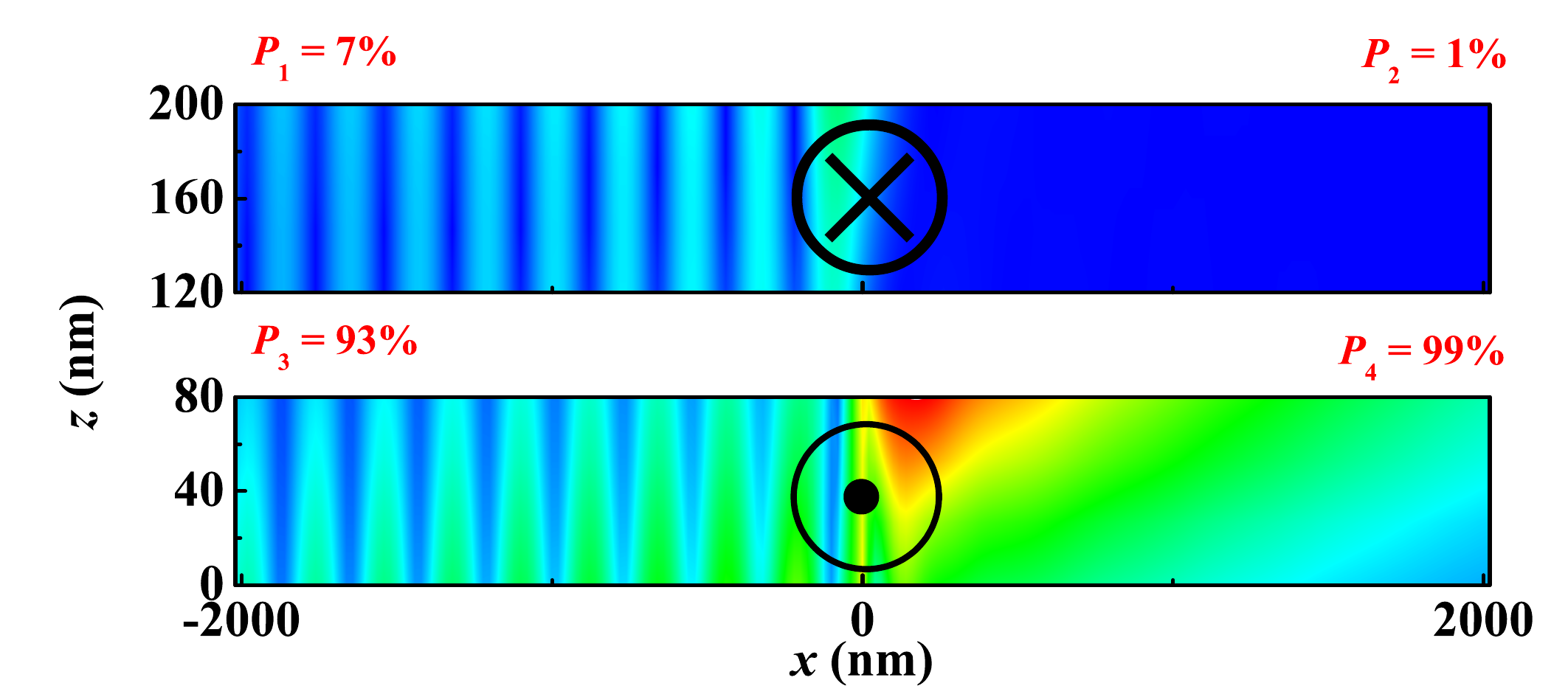}
    \caption{\label{waveguide-anti+ey-848} Spatial profiles of amplitudes of propagating SWs in two antiparallel films. The external electric field $ E_z = 0.34 $ MV/cm, and the frequency of the rf field is 4.24 GHz.}
\end{figure}

The external electric field $ E_z $ applied to both films shifts the SW dispersion (see Eqs. (\ref{coupled1}) and (\ref{coupled2})), as demonstrated in Fig. \ref{ana-ele-dispersion}.  Shifting of the dispersion relations changes the value of $ L $, leading to the electrically reconfigurable directional coupler . In Figs. \ref{waveguide-anti+ey-707} and \ref{waveguide-anti+ey-848}  the output $ P_i $ is plotted for the electric field $ E_z = 0.34 $ MV/cm, whereas in (Figs. \ref{waveguide-anti-707} and \ref{waveguide-anti-848}) the electric field is zero, and the value of the output $ P_i $ is different.
\begin{figure}
    \includegraphics[width=0.48\textwidth]{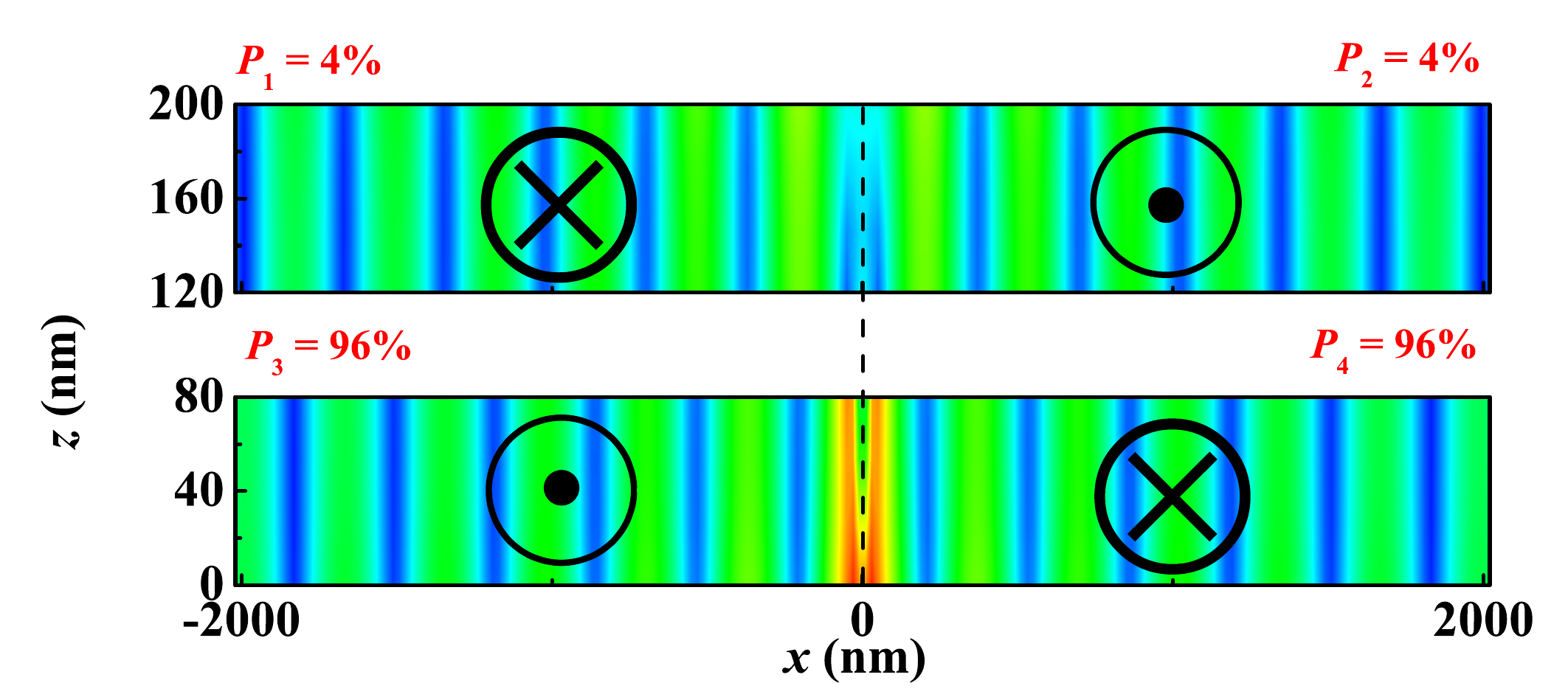}
    \caption{\label{antipal-domainwall-669} Spatial profiles of amplitudes of propagating SWs in two antiparallel films with domain walls. The frequency of the rf field is 3.34 GHz.}
\end{figure}

\begin{figure}
    \includegraphics[width=0.48\textwidth]{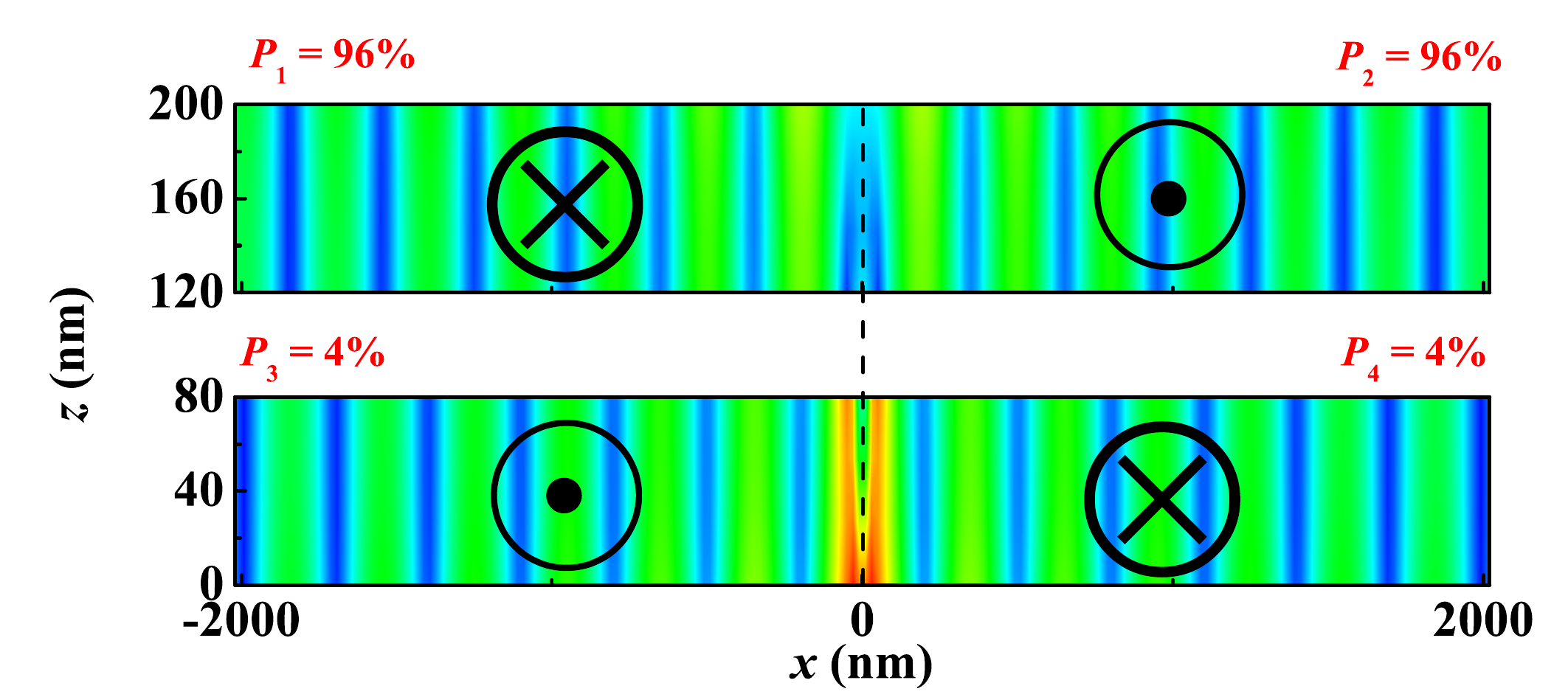}
    \caption{\label{antipal-domainwall-705} Spatial profiles of amplitudes of propagating SWs in two antiparallel films with domain walls. The frequency of rf field is equal to 3.52 GHz.}
\end{figure}

We note that the magnetization moments in four different regions (Fig. \ref{coupler-device}) near the output terminals accept the selective individual control, and this is a definite advantage of the functionality of proposed directional coupler. The control is  achievable  through the
current-induced spin-orbit torque and external magnetic field.  In Figs. \ref{antipal-domainwall-669} and \ref{antipal-domainwall-705}, we show amplitudes of propagating MSSWs  in the anti-parallel films with magnetic domain walls located near the center ($ x = 0 $).  SWs propagating in $ +x $ and $ -x $ directions share the same chirality, and thus we obtain symmetric outputs in anti-parallel films, i.e. $ P_1 = P_2 $ and $ P_3 = P_4 $.

For an experimental realization  we suggest the following schemes for exciting spin waves with  wave-vectors up to $ 0.04 $ nm$ ^{-1} $. The emergent net ferroelectric polarization coupled (through the magneto-electric coupling) to the external electric field mimics an effective DM term. Thus, the applied nonuniform electric field is equivalent to the nonuniform DM term and leads to a particular type of the torque termed as inhomogeneous electric torque \cite{prb064426}. The expression of the inhomogeneous electric torque is similar to the spin-transfer torque $ l_E \vec{m} \times (\vec{m} \times \vec{p}_E) $. The vector $ \vec{p}_E = \vec{x} \times \vec{e}_i $, $ \vec{e}_{i=x,y,z} $ points to the direction of the electric field, and the electric field gradient $ \partial_x E_i$ determines the coefficient $ l_E = - \gamma c_E \partial_x E_i / (\mu_0 M_s)$.  Therefore, the applied  oscillating inhomogeneous electric field $ E_y = 0 (x < 0) $ and $ E_y = E_1(t) (x > 0) $, with $ \vec{p_E} = \vec{z} $ at $ x = 0 $, leads to an oscillating electric torque, excites the magnetization oscillation at $x = 0$ and short-wavelength spin waves.  The  inhomogeneous electric field can be realized by a combination of  uniform electric field and a normal metallic cap layer shielding the region $ x < 0 $. Besides, via the microwave magnetic field from antennas, one can excite  large wavelength spin-waves in the region of a large internal effective magnetic field. The wavelength of the wave becomes smaller as soon as the wave crosses the region of the lower internal effective magnetic field. This method was proposed in Ref. [\mycite{KostylevDemidov}]. The different effective magnetic fields can be achieved by changing the geometry of the sample, attaching an exchange bias layer to the part of the magnetic layer, or covering the waveguide by the superconducting material (at the temperature below $T<T_{c}$).

\section{Switching of the static magnetization}

\begin{figure}
    \includegraphics[width=0.48\textwidth]{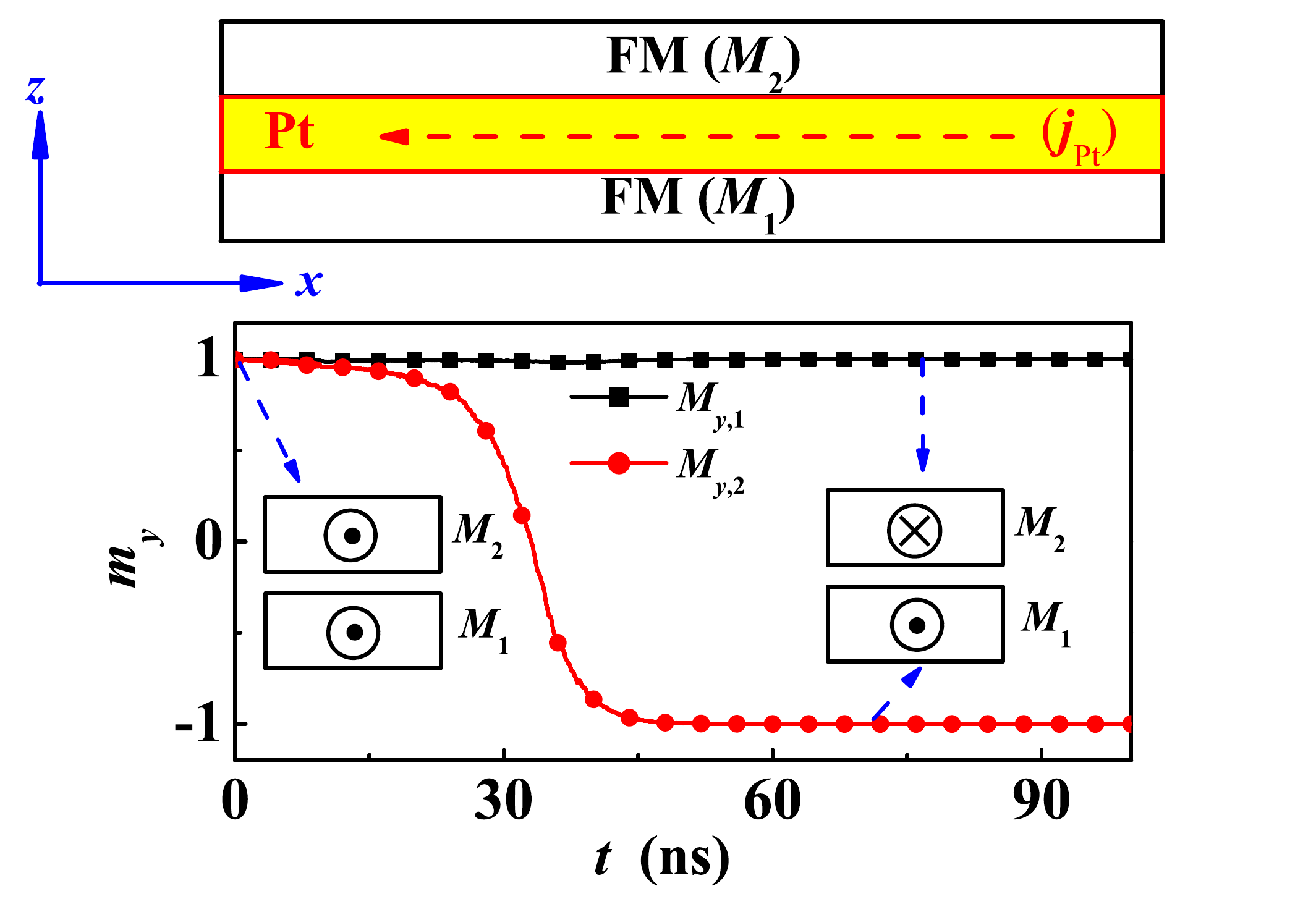}
    \caption{\label{anti-torque} Generation of the anti-parallel ground state. In the schematics (upper panel), a Pt spacer is placed between two dipolarly coupled films (FMs). Injecting an electronic current $ \vec{j}_{\mathrm{Pt}} $ in $ -\vec{x} $ direction, and applying opposite spin Hall torques on two films with parallel initial state, reverses the magnetization of one film ($ \vec{M}_2 $ in this figure) after dozens of nanoseconds. The switching process (time dependence of $ M_y $) is demonstrated in the lower panel.}
\end{figure}

\begin{figure}
    \includegraphics[width=0.48\textwidth]{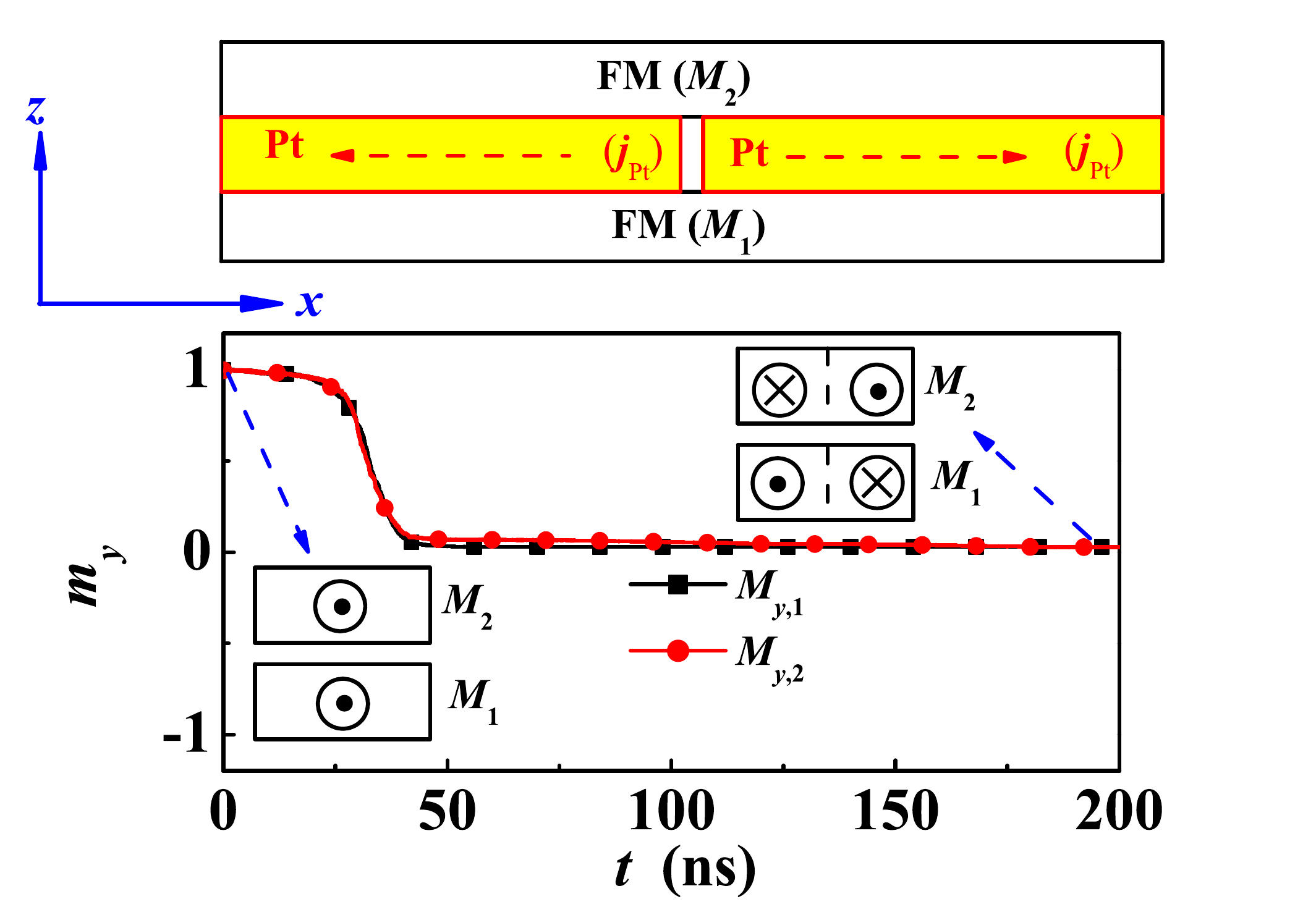}
    \caption{\label{domainwall-torque} Generation of the anti-parallel domain wall state. Two independent Pt spacer layers (see top panel) are embedded between two films (FMs). After injecting $ \vec{j}_{\mathrm{Pt}} $ with different directions in two Pt layers, the induced spin Hall torque generates the desired domain wall structure. The switching process (time dependence of $ M_y $) is demonstrated in the lower panel, and the gap between two Pt layers along $ x $ axis is 200 nm.}
\end{figure}

For practical realization, we suggest using the spin Hall torque for switching the magnetization in coupled films. As is shown in Fig. \ref{anti-torque}, the Pt spacer with electronic current $ \vec{j}_{\mathrm{Pt}} $ in $ -\vec{x} $ direction generates the spin Hall torques $ \gamma c_j \vec{M} \times \vec{\sigma} \times \vec{m}$. Here, $ \vec{\sigma} = \mp \vec{z} \times \vec{j}_{\mathrm{Pt}} $ represents the polarization direction, the coefficient $ c_j $ is proportional to the electronic current density in Pt, and $ -\vec{z} $ and $ + \vec{z} $ correspond to the bottom and the top layers. Also, we consider a small anisotropy constant $ K_y = 3500 $ J/m$^3$ (axis is along $ y $) in both films. Torques have opposite directions in layers and $ \vec{M}_{1,2} $ tends to be aligned along the $ \pm \vec{y} $ direction. For $ c_j = 790 $ A/m, switching of magnetizations induced by the spin Hall torque is shown in Fig. \ref{anti-torque}. In dozens of nanoseconds the state with parallel magnetizations is switched to the  anti-parallel state. After turning off the electric current, the system stays in anti-parallel state.  By applying a strong enough uniform magnetic field, one can easily switch the state back to the parallel state.

To generate the desired configuration of magnetic domain walls, we use the spin Hall torque, as demonstrated in Fig. \ref{domainwall-torque}. Two Pt layers with opposite electric currents generate opposite torques in two edges of the single film. The torques are also opposite in different films. The spin Hall torques with $ c_j = 790 $ A/m generate the particular structure of domain wall as studied here, e.g. in Fig. \ref{domainwall-torque}. The magnetic domain walls are stable after the electric current is turned off.
\section{Conclusions}

 Magnetostatic surface waves and surface waves propagating in coupled parallel wave guides are prototype examples of the simplest integrated magnonic circuits. The surface waves coupled through the dipole-dipole interaction periodically exchange energy. For a high fidelity of magnonic gates, the process of energy exchange should be well under control. In the present work, we proposed a particular type of surface wave magnonic gate controlled through an external static electric field. By varying the strength or direction of electric field, it is possible to shift selectively the chiral MSSW dispersion relation in coupled waveguide. The fact is used for manipulation of the propagation and transfer of the SWs in waveguides. Combining this effect, we propose an electrically reconfigurable nanoscale spin wave directional coupler, and its outputs can be selectively changed via controlling the electric field or magnetization. The adopted electric field is easy to realize and control, which is helpful for designing high fidelity surface wave magnonic devices.

 \section{Acknowledgment}
 We acknowledge financial support from the National Natural Science Foundation of China (No. 11704415, 11674400, 11374373), DFG through SFB 762 and SFB TRR227, and the Natural Science Foundation of Hunan Province of China (No. 2018JJ3629).

\end{document}